\documentclass[useAMS,usenatbib]{mn2e}

\usepackage{aecompl}
\usepackage[toc,page]{appendix}

\usepackage{graphicx}
\usepackage{amssymb}
\usepackage{amsmath}
\usepackage{aas_macros}
\usepackage{deluxetable}
\usepackage{lscape}
\usepackage{rotating}
\usepackage[symbol*]{footmisc}
\usepackage{gensymb}
\usepackage{tikz}



\title[GMF G214.5-1.8 as traced by CO]{GMF G214.5-1.8 as traced by CO: I - cloud-scale CO freeze-out as a result of a low cosmic-ray ionisation rate}

\author[S. D. Clarke et al.]{S. D. Clarke$^{1}$\thanks{E-mail: sclarke@asiaa.sinica.edu.tw}, 
V. A. Makeev$^{2,3}$, 
{\'A} S{\'a}nchez-Monge$^{4,5}$, 
G. M. Williams$^{6,7}$, 
Y.-W. Tang$^{1}$,\newauthor 
S. Walch$^{8}$, 
R. Higgins$^{8}$, 
P. C. N{\"u}rnberger$^{8}$ 
and S. Suri$^{9}$.
\\
$^{1}$Institute of Astronomy and Astrophysics, Academia Sinica, No. 1, Section 4, Roosevelt Road, Taipei 10617, Taiwan\\
$^{2}$Moscow Institute of Physics and Technology, Institutsky per. 9, Dolgoprudny 141700, Russia\\
$^{3}$Lebedev Physical Institute of the Russian Academy of Sciences, Leninsky prospekt 53, 119991 Moscow, Russia\\
$^4$Institut de Ci\`{e}ncies de l'Espai (ICE, CSIC), Can Magrans s/n, E-08193, Bellaterra, Barcelona, Spain\\
$^5$Institut d'Estudis Espacials de Catalunya (IEEC), Barcelona, Spain\\
$^6$School of Physics and Astronomy, University of Leeds, Leeds, LS2 9JT, UK\\
$^7$Department of Physics, Aberystwyth University, Ceredigion, Cymru, SY23 3BZ, UK\\
$^{8}$I. Physikalisches Institut, Universit{\"a}t zu K{\"o}ln, Z{\"u}lpicher Str. 77, D-50937 K{\"o}ln, Germany \\
$^9$University of Vienna, Department of Astrophysics, T{\"u}rkenschanzstrasse 17, A-1180 Vienna, Austria}

\newcommand{\Su}{_{_{\odot}}}
\newcommand{\co}{^{12}\textrm{CO}}
\newcommand{\coc}{^{13}\textrm{CO}}
\newcommand{\coo}{\textrm{C}^{18}\textrm{O}}
\begin{document}

\date{}

\pagerange{\pageref{firstpage}--\pageref{lastpage}} \pubyear{2002}

\maketitle

\label{firstpage}

\begin{abstract}
We present an analysis of the outer Galaxy giant molecular filament (GMF) G214.5-1.8 (G214.5) using IRAM 30m data of $\co$, $\coc$ and $\coo$. We find that the $\co$ (1-0) and (2-1) derived excitation temperatures are near identical and are very low, with a median of 8.2 K, showing that the gas is extremely cold across the whole cloud. Investigating the abundance of $\coc$ across G214.5, we find that there is a significantly lower abundance along the entire 13 pc spine of the filament, extending out to a radius of $\sim 0.8$ pc, corresponding to $A_v \gtrsim 2$ mag and $T_{dust} \lesssim 13.5$ K. Due to this, we attribute the decrease in abundance to CO freeze-out, making G214.5 the largest scale example of freeze-out yet. We construct an axisymmetric model of G214.5's $\coc$ volume density considering freeze-out and find that to reproduce the observed profile significant depletion is required beginning at low volume densities, $n\gtrsim2000$ cm$^{-3}$. Freeze-out at this low number density is possible only if the cosmic-ray ionisation rate is $\sim 1.9 \times 10^{-18}$ s$^{-1}$, an order of magnitude below the typical value. Using timescale arguments, we posit that such a low ionisation rate may lead to ambipolar diffusion being an important physical process along G214.5's entire spine. We suggest that if low cosmic-ray ionisation rates are more common in the outer Galaxy, and other quiescent regions, cloud-scale CO freeze-out occurring at low column and number densities may also be more prevalent, having consequences for CO observations and their interpretation.    
\end{abstract}


\section{Introduction}\label{SEC:INTRO}%
The interplay of gravity, turbulence and magnetic fields in the interstellar medium (ISM) leads to a complex hierarchy of structures of various morphologies across a range of spatial scales \citep[see][for a recent review]{Pin23}. One recent finding has been that the cold, dense molecular ISM is predominately arranged into filaments, and that these filaments are linked to the star formation process \citep{And14}. This has lead to a surge of observational and theoretical works to better understand the role of these filamentary structures in the last decade \citep[e.g.][]{Arz13,Smi14,Cla15,Kon15,Cla16,Cox16,Mar16,Cla17,Wil18,Cla18,How19,Suri19,Bon20a,Bon20b,Cla20,Abe21,Li22,Wang22,Bon23}.

\begin{figure*}
\centering
\includegraphics[width=0.9\linewidth, trim={0.75cm 0.75cm 0.75cm 0cm},clip]{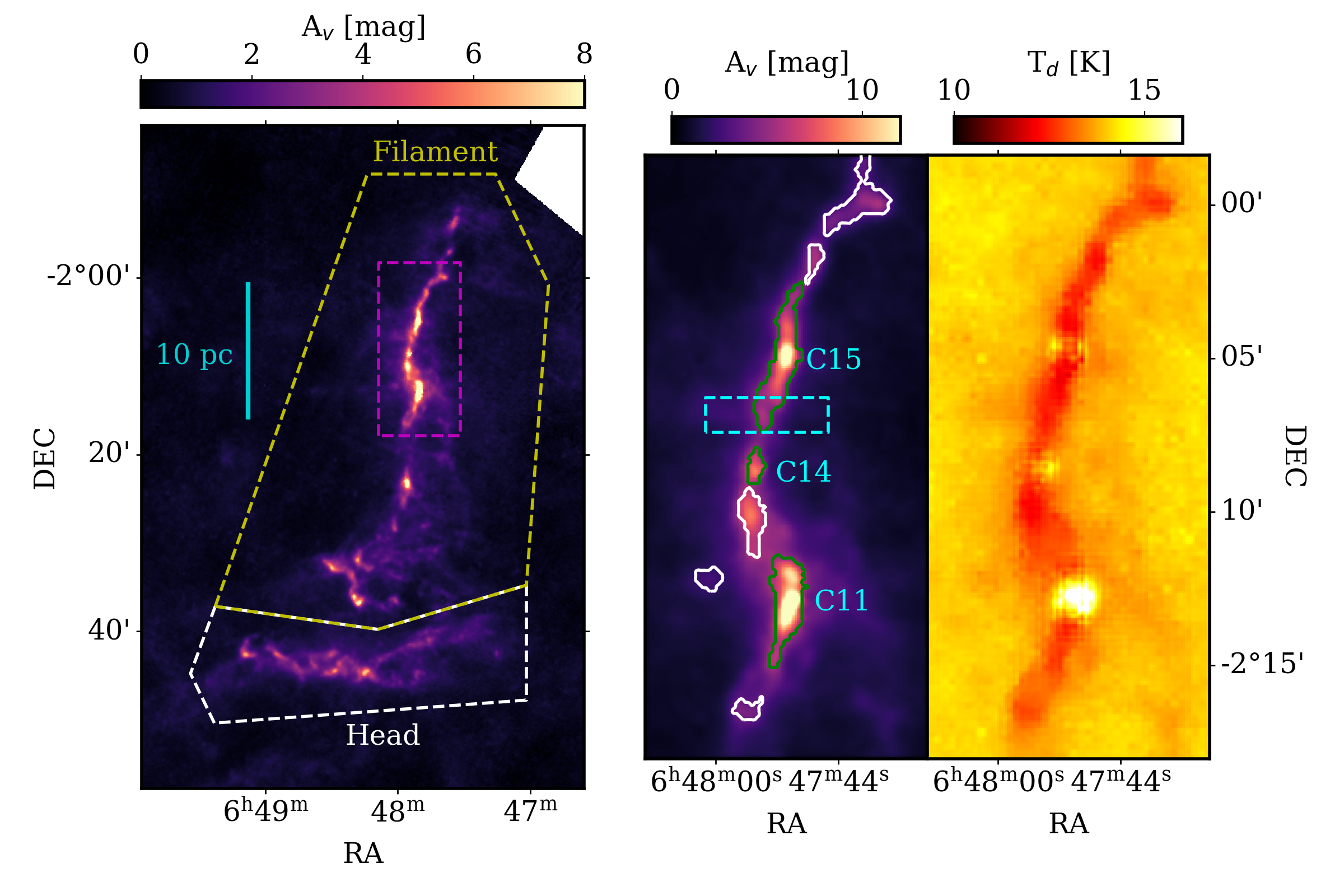}
\caption{(Left) A 18.2'' \textit{Herschel} column density map of the entire G214.5 giant molecular filament, with the \textit{main filament} and \textit{head structure} enclosed by the yellow and white dashed polygons respectively. The magenta, dashed rectangle shows the region mapped using the IRAM 30m telescope. The cyan scale shows the apparent size of 10 pc at the distance of 2.3 kpc. (Middle) The \textit{Herschel} column density map of the region enclosed by the magenta rectangle when smoothed to 23.5'' with contours denoting the clumps identified in \citet{Cla23}. Green and white contours show protostellar and starless clumps respectively. The cyan, dashed rectangle denotes the region mapped using APEX. The 3 protostellar clumps of interest, C11, C14 and C15, are also labelled. (Right) The 36'' \textit{Herschel} dust temperature map of the region enclosed by the magenta rectangle.}
\label{fig::HerCD}
\end{figure*}  

Large-scale Galactic surveys have been a highly important driver in our understanding of the complex web of structures in the ISM and provided a manner to investigate the nature of molecular clouds in a statistical sense \citep[e.g.][]{Hey98,Dame01,Jac06,Sch09,Mol10,Bur13,Dem13,Beu16,Rig16,Sch17,Ume17,Su19}. A common molecular line tracer used to study the kinematics of molecular clouds in these surveys is $\co$ and its isotopologues $\coc$ and $\coo$. This is due to their relatively high abundance and brightness making them ideal when observing large areas. Numerous works have studied what gas is traced by CO, relating emission to physical quantities such as density and mass, as well as examining the chemical/radiative effects which impact the CO abundance and the ratio between isotopologues \citep[e.g.][]{VanDis88,Glo10,Bol13,Szu14,Nis15,Szu16,Pen17,Pen18,Are18,Are19,ClaGlo19,Rig19,Rou21,Bor22,Eba22}.

One important discovery from Galactic surveys is the presence of giant molecular filaments (GMFs) which are typically defined as velocity coherent, large-scale ($\gtrsim10$ pc), high aspect ratio, massive clouds \citep{Jac10,Goo14,Rag14,Wang15,Abr16,Zuc18,Zha19,Wang20,Col21,Ge23}. It is currently not known how GMFs form, or if there is a universal mechanism, but it is thought that the majority form from a combination of shear and the Galactic potential leading to the elongation of existing clouds \citep{Dob06,Dob11,Dua16,Dua17,Smi20}. As such, these large structures are potentially shaped by their Galactic environment whilst also being regions of extended star formation, making GMFs an excellent place to investigate how large-scale ISM processes affect the star formation process. One such GMF is the outer Galaxy cloud G214.5-1.8, which is the focus of this paper.

\subsection{The giant molecular filament G214.5-1.8}\label{SSEC:G214}%
G214.5-1.8 (hereafter G214.5) is a giant molecular filament associated with the larger Maddalena's cloud and lies at a distance of 2301$\pm150$ pc \citep{MadTha85,Lee91,Yan19}. It has recently been the focus of a \textit{Herschel} study by \citet{Cla23}. They find that G214.5 has a mass of $\sim 16,000$ M$\Su$ and a length of approximately 35 pc from north to south, making it comparable to other GMFs studied \citep{Zuc18,Zha19,Col21}. However, they also find that G214.5 is highly quiescent for its mass and size, hosting very few 70 $\mu$m sources. Due to this, as well as the paucity of dense gas, they posit that G214.5 is a young GMF which may not have began forming stars in earnest.

As can be seen in the column density map\footnote{Column density data taken from \citet{Cla23} where the authors have already subtracted the unrelated line-of-sight contribution to the column density.} in figure \ref{fig::HerCD}, G214.5 has a unique morphology which consists of two distinct subregions, a north to south thin filament and an east to west flocculent structure; \citet{Cla23} name these the \textit{main filament} and the \textit{head structure} respectively. These two subregions are velocity coherent and a smooth velocity gradient exists across the cloud from $\sim22 - 32$ km/s \citep{Lee91,Yan19}. While the average properties of the two subregions are similar, \citet{Cla23} find that the vast majority of the dense gas and protostellar clumps are located in the \textit{main filament} and the \textit{head structure} is more diffuse.

A possible explanation for the unusual morphology of G214.5 is that of an interaction with a neighbouring HI superbubble. This is suggested by \citet{Cla23} due to G214.5 lying on the edge of the superbubble GSH214.0+00.0+017.5, being coincident in both the plane of the sky and line-of-sight velocity. Further, \citet{Cla23} find evidence of an interaction in the strongly asymmetric radial profiles along the lower section of the \textit{main filament}. These profiles are highly reminiscent of those found in filaments known to be impacted by external feedback \citep[e.g.][]{Per12,Zav20}. Informed by the simulations of \citet{Gol20}, \citet{Cla23} suggest that a mixture of compression and erosion by the superbubble front may lead to the formation of the narrow, dense \textit{main filament} and the flocculent, extended \textit{head structure}. Thus, G214.5 is an excellent candidate for studying the evolution of a young, quiescent GMF in the context of the bubble-dominated ISM \citep{Inu15,Wal15,Pin23}.

The aim of this paper is to use IRAM 30m observations of the isotopologues $\co$, $\coc$ and $\coo$ to study the physical properties of G214.5 and better understand what CO observations may tell us of young GMFs in quiescent regions. The paper is structured as follows: in section \ref{SEC:OBS} we outline the details of the IRAM 30m, APEX and SOFIA observations; in section \ref{SEC:RES} we present moment maps and discuss general features seen in the observations; in section \ref{SEC:NCO} we detail the method used to calculate CO column densities, study the $\co$ excitation temperatures, present $\coc$ and $\coo$ abundances and determine the isotopologue ratio X$_{13/18}$ = N$_{\coc}$/N$_{\coo}$; in section \ref{SEC:CON} we summarise our results and explore their implications for CO observations in quiescent regions such as the outer Galaxy.

\section{Observations}\label{SEC:OBS}%

The IRAM 30m telescope was used in the on-the-fly (OTF) mapping mode between September 4th and 10th 2019 under project 009-19 (PI: S. Clarke). The mapped region is approximately 8' $\times$ 19' and is shown as the magenta, dashed rectangle in the left panel of figure \ref{fig::HerCD}. To ensure reasonable observing times per OTF map, the region was divided into 80'' $\times$ 80'' squares with a total of 73 square maps observed. Each square was observed twice, once using right ascension aligned scans and once using declination aligned scans, to minimise mapping artefacts. Position-switching was employed using the OFF position at (RA,DEC) = (06:48:37.7,-02:08:48.7). The accuracy of the pointing was checked every 1-2 hours.

Observations were performed using the EMIR heterodyne receiver tuned to observe the E090 band at 107.4-115.4 GHz and the E230 band at 228.0-236.0 GHz simultaneously. This allowed coverage of the $\co$, $\coc$ and $\coo$ (1-0) lines at 115.27 GHz, 110.20 GHz and 109.78 GHz respectively and the $\co$ (2-1) line at 230.54 GHz. This results in spatial resolutions of $\sim22$'' and $\sim11$'' for the two bands respectively. The FTS200 spectrometer was set as the backend leading to a spectral resolution of 195 kHz, corresponding to $\sim0.6$ km/s velocity resolution for the E090 band and $\sim0.3$ km/s for the E230 band. The VESPA backend was also used to cover the $\coc$ and $\coo$ (1-0) lines to provide 0.2 km/s resolution data; these data are not considered here but will be used in a companion work focusing on the kinematics of G214.5 (Clarke et al. in prep.).

The data was reduced using the CLASS and GREG programs from the GILDAS software package\footnote{https://www.iram.fr/IRAMFR/GILDAS}. To convert from antenna temperature (T$_a^*$) to main beam temperature (T$_{\rm mb}$), we used the equation T$_{\rm mb}$ = (F$_{\rm eff}$/B$_{\rm eff}$)T$_a^*$, where F$_{\rm eff}$ is the forward efficiency and B$_{\rm eff}$ is the main beam efficiency; for the E090 band we used B$_{\rm eff}$ = 0.78 and F$_{\rm eff}$ = 0.94, and for the E230 band we used B$_{\rm eff}$ = 0.59 and F$_{\rm eff}$ = 0.92\footnote{Values taken from the table provided by IRAM, https://publicwiki.iram.es/Iram30mEfficiencies}. All data were binned to a common 0.6 km/s velocity resolution and mapped to 4'' pixel size. Due to the varying spatial resolutions across the lines we smoothed all data to a common 23.5'' resolution. The resulting noise maps can be seen in figure \ref{fig::app_noise} in appendix \ref{APP:ERROR}; the mean noise levels are 206 mK, 96 mK and 96 mK for the $\co$, $\coc$ and $\coo$ (1-0) lines respectively, and 217 mK for the $\co$ (2-1) line.

Additional observations were taken with the APEX telescope using the LASMA instrument between the 11th and 18th of October 2019 under project O-0103.F-9318A-2019 (PI: S. Clarke); only the single central pixel of the LASMA instrument was used. A region of approximately 250'' by 60'' was mapped which lies across the spine of G214.5 away from the clumps, see the cyan, dashed rectangle in the middle panel of figure \ref{fig::HerCD}, using the OTF mode. The same OFF position as that used for the IRAM 30m observations was used. These observations allowed the simultaneous observations of the $\co$ and $\coc$ (3-2) lines at 345.80 GHz and 330.59 GHz respectively. The beam size at these frequencies is $\sim19$''. The data were binned onto a 9.5'' pixel size map with a velocity resolution of 0.25 km/s, and a main beam efficiency of 0.74 was used to convert from antenna temperature to main beam temperature. The mean noise levels are 34 mK and 41 mK for the $\co$ and $\coc$ lines respectively. 

Observations were also taken using the upGREAT instrument \citep{Ris18} on the Stratospheric Observatory for Far-Infrared Astronomy (SOFIA) on the 11th of March of 2020 under project 83\_0723 (PI: S. Clarke). These observations consisted of 3 pointings towards the clumps C11, C14 and C15 in G214.5 with positions (RA,DEC) = (06:47:50.2,-02:12:53), (06:47:55,-02:08:38) and (06:47:51,-02:05:01) respectively. Each pointing contains 7 pixels; their locations are shown in figure \ref{fig::app_CII} in appendix \ref{APP:CII}. The OFF position was the same as that used for the IRAM 30m and APEX observations. Main beam efficiencies for each pixel were determined through observations of Mars, resulting in efficiencies between 0.63 and 0.69. The low-frequency and high-frequency arrays, upGREAT/LFA and upGREAT/HFA, were used to simultaneously observe the 157.7 $\mu$m [CII] and 63.2 $\mu$m [OI] fine structure lines respectively; however, in this work we consider only the [CII] data. The beam size at 157.7 $\mu$m is 14.1'', and the data was smoothed to a velocity resolution of 0.38 km/s. The achieved noise level for the [CII] spectra are between 0.2 and 0.5 K depending on the pixel and the pointing, with a mean of 0.32 K.

\section{Results}\label{SEC:RES}%

\begin{figure*}
\centering
\includegraphics[width=0.85\linewidth, trim={0cm 2.1cm 1.55cm 1.55cm},clip]{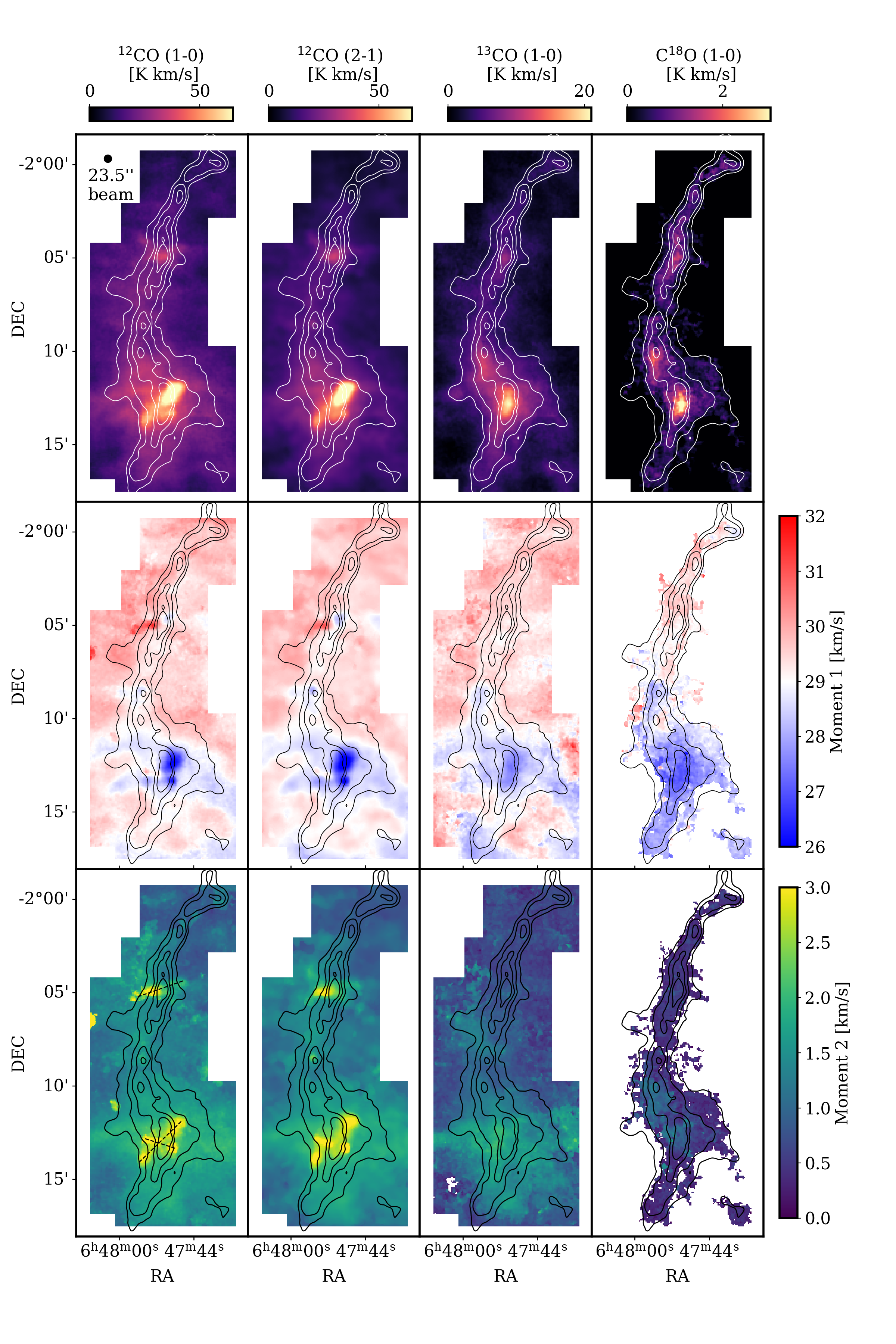}
\caption{Moment maps for, from left to right, $^{12}$CO (1-0), $^{12}$CO (2-1), $^{13}$CO (1-0), C$^{18}$O (1-0). The top row shows the moment zero, i.e. integrated intensity, the middle row shows the moment one, i.e. velocity centroid, and the bottom row shows the moment two, i.e. velocity dispersion. The contours denote total column density at an A$_v$ of 2, 3, 5 and 8 magnitudes, taken from the Herschel column density maps of \citet{Cla23}. The top left panel shows the 23.5'' beam size as a black circle. The black, dashed lines in the moment 2 panel for $\co$ (1-0) denote the identified locations of outflow lobes.}
\label{fig::mom}
\end{figure*}

We use a moment masking technique to create moment zero, one and two maps from the $\co$ (1-0), $\co$ (2-1), $\coc$ (1-0) and $\coo$ (1-0) lines. We use the moment masking technique presented by \citet{Dame11} as it is shown to produce `cleaner' and more accurate moment maps compared to the typical clipping technique. We modify the technique slightly to aid in detecting weaker emission. \citet{Dame11} unmask voxels which are within the smoothing kernel of voxels which themselves are unmasked by some threshold, $T_C$ (step vii in their section 3). Here we use an iterative method which unmasks voxels that neighbour previously unmasked voxels as long as they have a value above some threshold, $T_L$ where $T_L < T_C$. We use the $\textsc{make\_moments}$ function in the latest version of $\textsc{BTS}$ for this purpose \citep{Cla18} and full details of the moment masking method can be found in its documentation\footnote{https://github.com/SeamusClarke/BTS}. After testing various values of $T_C$ and $T_L$, we find that 8$\sigma$ and 3$\sigma$ respectively produce moment maps which are least affected by noise while capturing weaker emission. The resulting moment zero, one and two maps are shown in figure \ref{fig::mom}. The uncertainty on the moment zero values can be found in the online supplementary materials.

One can see from the moment zeros maps that $\co$ and $\coc$ are detected across the entire mapped region, while $\coo$ is only detected in the dense filament, i.e. approximately A$_v > 2$ mag. Further, while the spatial distribution of $\coc$ and $\coo$ are similar to the column density derived from dust emission, the $\co$ lines show a flatter spatial distribution with a large bright bar protruding from the southern massive clump, C11, and a less prominent x-shaped feature towards a the northern clump, C15. These are the result of large-scale outflows from these clumps, as evidenced by the associated large red-blue velocities in the moment one map and the localised, elevated dispersion in the moment two maps (outflows are identified as black, dashed lines in the $\co$ (1-0) moment 2 panel of figure \ref{fig::mom}). A bipolar outflow from clump C11 has been previously reported in CS \citep{Lar99}, but examination of the moment one and moment two maps suggests two potential outflows with position angles of 45.6$\degree$ and 109.5$\degree$ with respect to north. These outflow lobes have approximate lengths of 96'' and 54'' respectively, which at a distance of 2.3 kpc results in sizes of 1.07 pc and 0.60 pc. The two outflows do not intersect at the centre of clump C11; however, higher resolution observations are needed to determine if this is accurate or a result of low spatial resolution. The x-shaped feature from the northern clump C15 has a clear red-blue gradient, indicative of an outflow, along only one of its `strokes'. This outflow has not previously been reported and has a position angle of 71.8$\degree$, making it approximately perpendicular to the filament spine, and a length of 78'' or 0.87 pc. \citet{Cla23} note that C15 contains two 70$\mu$m sources, S8 and S9; on close inspection it appears that the outflow from this clump originates from the dimmer S8 source. Investigating the moment 1 velocity from the $\co$ (1-0) map along these outflow axes, we find the difference between the most blue-shifted and most red-shifted velocities, $\Delta V$. We may define a dynamical time of these outflows as $2L/\Delta V$, resulting in values of 0.58 Myr and 0.31 Myr for the two outflows from C11 and 0.60 Myr for the outflow from C15. These dynamical timescales are only approximate as the large beam size does not allow one to resolve the internal structures of the outflows, but their low values do support the youth of G214.5 proposed by \citet{Cla23}.

Examination of the moment one maps show that the filament is velocity coherent across its length and has a velocity gradient from north to south, seen in all three isotopologues. The line-of-sight velocity varies from $\sim 27$ km/s in the south to $\sim$ 30 km/s in the north. The velocity dispersion, $\sigma_v$, of the $\coc$ and $\coo$ lines is low within the filament having values of $0.5-2$ km/s, with higher values located in the south around the C11 clump, potentially connected to the outflows. Further analysis of the kinematics of G214.5 is left to a companion paper (Clarke et al. in prep.). 

\citet{Cla23} note that their estimate of the mass of the entire G214.5 GMF using \textit{Herschel} data is approximately half that of the estimate from \citet{Yan19}, who use the canonical CO-to-H$_2$ conversion factor, $X_{\rm CO} = 2 \times 10^{20}$ cm$^{-2}$ (K km/s)$^{-1}$ \citep{Bol13}. For our mapped region we calculate the average $X_{\rm CO}$ factor, defined as $\sum N_{\rm H_2} / \sum I_{\rm CO}$, where $I_{\rm CO}$ is the moment zero of $\co$, to be $1.12 \times 10^{20}$ cm$^{-2}$ (K km/s)$^{-1}$. If this value held across the whole cloud then it would bring the \textit{Herschel} derived mass estimate of G214.5 into much closer agreement with the CO derived mass estimate by \citet{Yan19}. Further, low $X_{\rm CO}$ values ($\sim 1 \times 10^{20}$ cm$^{-2}$ (K km/s)$^{-1}$) are expected in young molecular clouds in which wide-spread star formation has yet to occur, due to the time delay between H$_2$ and CO formation \citep{Bor22}. Thus, the low average $X_{\rm CO}$ factor calculated here is in agreement with the potential youth of G214.5 proposed by \citet{Cla23}.

\subsection{SOFIA [CII] spectra}\label{SSEC:CII}%
It has previously been shown that [CII] may be used as a tracer of the FUV field due to it being produced by the ionisation of neutral carbon by UV photons above an energy of 11.2 eV \citep[e.g.][]{TieHol85}. Out of the 21 individual SOFIA spectra, 20 spectra show no detections. As such, we follow the method of \citet{Bon20b}, and use the non-detection of [CII] to place upper limits on G214.5's external FUV-field; details may be found in appendix \ref{APP:CII}. Using this method we find that the ambient FUV-field has a strength less than $\sim1.1-2.5$ G$_0$. Such a weak external radiation field is consistent with the low dust temperatures and the quiescent environment found by \citet{Cla23}.  

As mentioned above, 1 pixel shows significant detection of [CII], with a peak intensity of $\sim1.1$ K $\approx 5\sigma$ (see figure \ref{fig::app_detect}). This is the central pixel towards clump C11, i.e. the most massive clump in G214.5, $\sim230$ M$\Su$. With the lack of detections at all other positions, it is likely that the [CII] emission at this location is not from external radiation, but is internal to the clump. This may be the first evidence that clump C11 may harbour an intermediate/high-mass protostar as previous searches using H$_2$O and class II methanol masers have found no such sign \citep{Sly99,Fur03,Sun07}, as well as the clump being found to be radio-quiet at 5 GHz with the VLA \citep{Urq09}. A more detailed analysis of the [CII] detection will be presented in a future work focusing on clump C11. 

\section{CO column density and abundances}\label{SEC:NCO}%
In this section we use the assumption of local thermodynamic equilibrium (LTE) and a beam-filling factor of 1 to determine the column density of $\coc$ and $\coo$ without the assumption of canonical isotopologue ratios relative to $\co$ \citep[see e.g.][]{ManShi15}. Using this method, one determines the $\co$ excitation temperature and, assuming it is equal to that of $\coc$ and $\coo$, one may use it to calculate column densities of the two species. Note, this method does not allow an estimate of the $\co$ column density, but does allow one to determine the isotopologue abundance ratio $^{13}$CO over C$^{18}$O, X$_{13/18}$, which has been shown to vary across clouds \citep[e.g.][]{Are18,Are19,Lin21,Rou21}.  

To calculate uncertainties on the quantities of interest in this section we use a Monte Carlo method which takes the noise of each individual spectrum and propagates it forwards. This is fully discussed in appendix \ref{APP:ERROR} and maps of the uncertainty in quantities are shown in the online supplementary material. This method results in 10,000 instances of each quantity calculated. When reporting the quantity of interest (e.g. $\coc$ column density) we report the mean of this distribution and the uncertainty is taken as the distribution's standard deviation unless otherwise stated. Note that throughout this paper the notation $A^{+b}_{-c}$ is used to denote the median (A) and interquartile range (-c,+b) of a quantity, where the range is over all the pixels; therefore it is not a measure of uncertainty but of variation. Uncertainties will be stated as such or use the $\pm$ notation.

\subsection{$\co$ excitation temperature}\label{SSEC:TEX}%
The excitation temperature of $\co$ may be estimated under the assumption that emission is optically thick using the equation:
\begin{equation}
T_{\rm ex} = \frac{T_{ul}}{\ln{\left(1 + T_{ul}\left[T_R + \frac{T_{ul}}{e^{T_{ul}/T_{bg}}-1}\right]^{-1} \right)}},
\label{eq::tex}
\end{equation}
where $T_{ul}$ is the energy gap of the transition in Kelvin, $T_{bg}$ is the background temperature taken to be the CMB temperature of 2.7 K, and $T_R$ is the radiation temperature taken as the peak temperature of each spectrum. We may use equation \ref{eq::tex} for both the (1-0) and (2-1) transitions of $\co$ due to the high optical depth of both lines; for the (1-0) transition $T_{ul}$ is set to 5.53 K and for the (2-1) transition $T_{ul}$ is set to 11.07 K. Note, if the $\co$ emission suffers from strong self-absorption features this method produces reduced estimates of the excitation temperature; however, inspection of even the brightest spectra show no such absorption features so we believe this effect has minimal impact on the excitation temperature estimates. 

\begin{figure}
\centering
\includegraphics[width=0.92\linewidth, trim={0cm 1.5cm 2.2cm 0cm},clip]{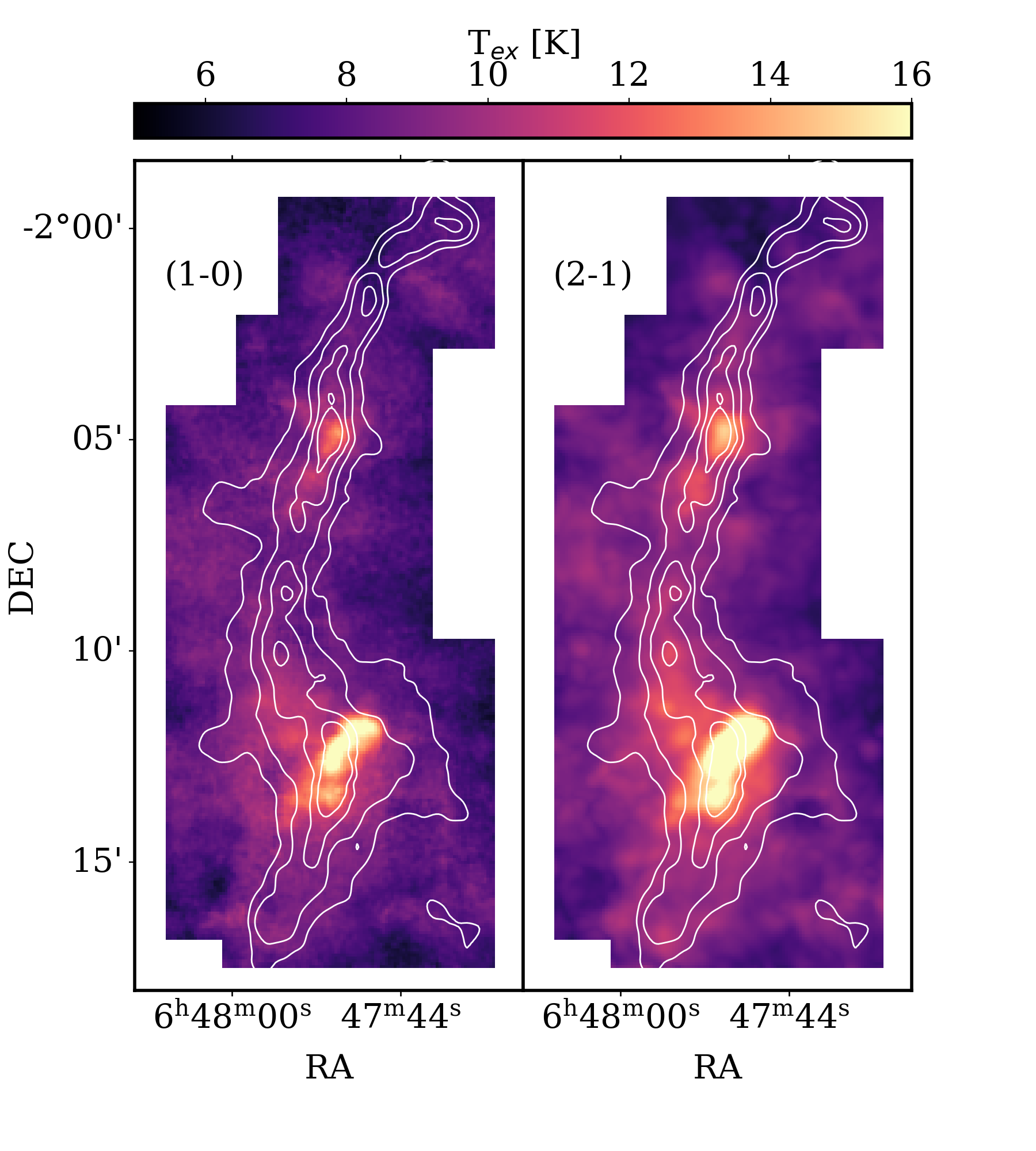}
\caption{A map of the excitation temperature calculated using the $^{12}$CO (1-0) and (2-1) lines under the assumption they are both optically thick (see equation \ref{eq::tex}). The contours denote total column density at an A$_v$ of 2, 3, 5 and 8 magnitudes.}
\label{fig::Tex}
\end{figure}  

Figure \ref{fig::Tex} shows the excitation temperature maps for both transitions of $\co$. One sees that the (1-0) and (2-1) transitions produce near identical maps of the excitation temperature, and that the measured excitation temperatures are low. The fact that the two transitions produce such similar excitation temperatures (the median of the ratio $T_{\rm ex, (2-1)} / T_{\rm ex, (1-0)}$=1.07$^{+0.04}_{-0.04}$) shows that the assumption that the gas is thermalised is valid and that the population levels may be well described by a single temperature. The low excitation temperature (median values over the whole map of 8.2$^{+0.7}_{-0.6}$ K and 8.7$^{+0.9}_{-0.6}$ K from the (1-0) and (2-1) lines respectively) is in agreement with that first reported for G214.5 by \citet{Lee94} which found excitation temperatures below 7.3 K to be common across G214.5 and the neighbouring Maddalena's cloud. The median error on the (1-0) line derived excitation temperature is 0.21 K, and as a percentage of the excitation temperature, the median error is 2.6$\%$, i.e. the excitation temperature is well constrained. Such low excitation temperatures are atypical and other molecular clouds (e.g. Serpens North, Ophiuchus and Orion B) show $^{12}$CO excitation temperatures lying in the range of 10-50 K \citep{Gra10,Whi15,Rou21}. There is some variation in the excitation temperature across the map in figure \ref{fig::Tex}, with a slight gradient of $\sim$0.5 K / 10$^{21}$ cm$^{-2}$; however, the results presented in the following sections are not affected by this variation as calculations using a fixed excitation temperature of 8.2 K (the median $\co$ (1-0) derived temperature) across the map yield the same results, as shown in the online supplementary material.

Low excitation temperatures may result from sub-thermal excitation \citep[see][]{Gol08,Pen17}, rather than low kinetic temperatures. However, as mentioned above, the near identical excitation temperatures derived from the (1-0) and (2-1) transitions show that the gas is thermalised and as such the excitation temperature is close to the gas kinetic temperature. This would mean that the CO bright gas in G214.5 is unusually cold. \citet{Cla23} show that the dust temperatures are also low, with values of around 12.5 K in the spine and 13.5-14.0 K at larger radii (see figure \ref{fig::HerCD}), making it colder than any previously studied inner Galaxy GMF. As there are few internal heating sources, all of which are localised to their immediate surroundings, i.e. the protostellar clumps C11, C14 and C15, no nearby OB stars/clusters \citep{Lee94,Wri23} and the [CII] non-detections constrain the ambient UV field to values below $\sim1.1$G$_0$, it is likely that the dominant heating mechanism is that due to cosmic rays. Thus, the low gas and dust temperatures indicate a decrease in the cosmic ray flux in G214.5's vicinity, in agreement with gamma-ray studies showing such a decrease with galactocentric radius \citep[e.g.][]{Yang16}.

It is known that there may exist differences in the excitation temperatures of the different CO isotopologues due to opacity and/or kinetic temperature gradients along the line-of-sight \citep[e.g. see][who find $\co$ excitation temperatures in general 70$\%$ greater than that for $\coc$ in Orion B]{Rou21}. However, an assumption we make in the following section when calculating the column densities of $\coc$ and $\coo$ is that their excitation temperatures are equal to that of $\co$. We find that this assumption is valid, within uncertainties, in the region in which the $\coc$ (3-2) line was observed with APEX. This is done by using non-LTE RADEX calculations to model the brightness and ratio of the $\coc$ (3-2) and (1-0) transitions; full details can be found in appendix \ref{APP:RADEX}. As the region mapped by APEX is a typical off-clump area across the spine, we believe that our finding that the $\co$ and $\coc$ excitation temperatures are approximately equal is likely to hold across the rest of the map; however, this may not be valid in the outflow effected regions close to clumps C11 and C15, where elevated $\co$ excitation temperatures are found. 

\subsection{$\coc$ and $\coo$ column densities and abundances}\label{SSEC:NCO}%

\begin{figure}
\centering
\includegraphics[width=0.92\linewidth, trim={0cm 1.5cm 2.2cm 0cm},clip]{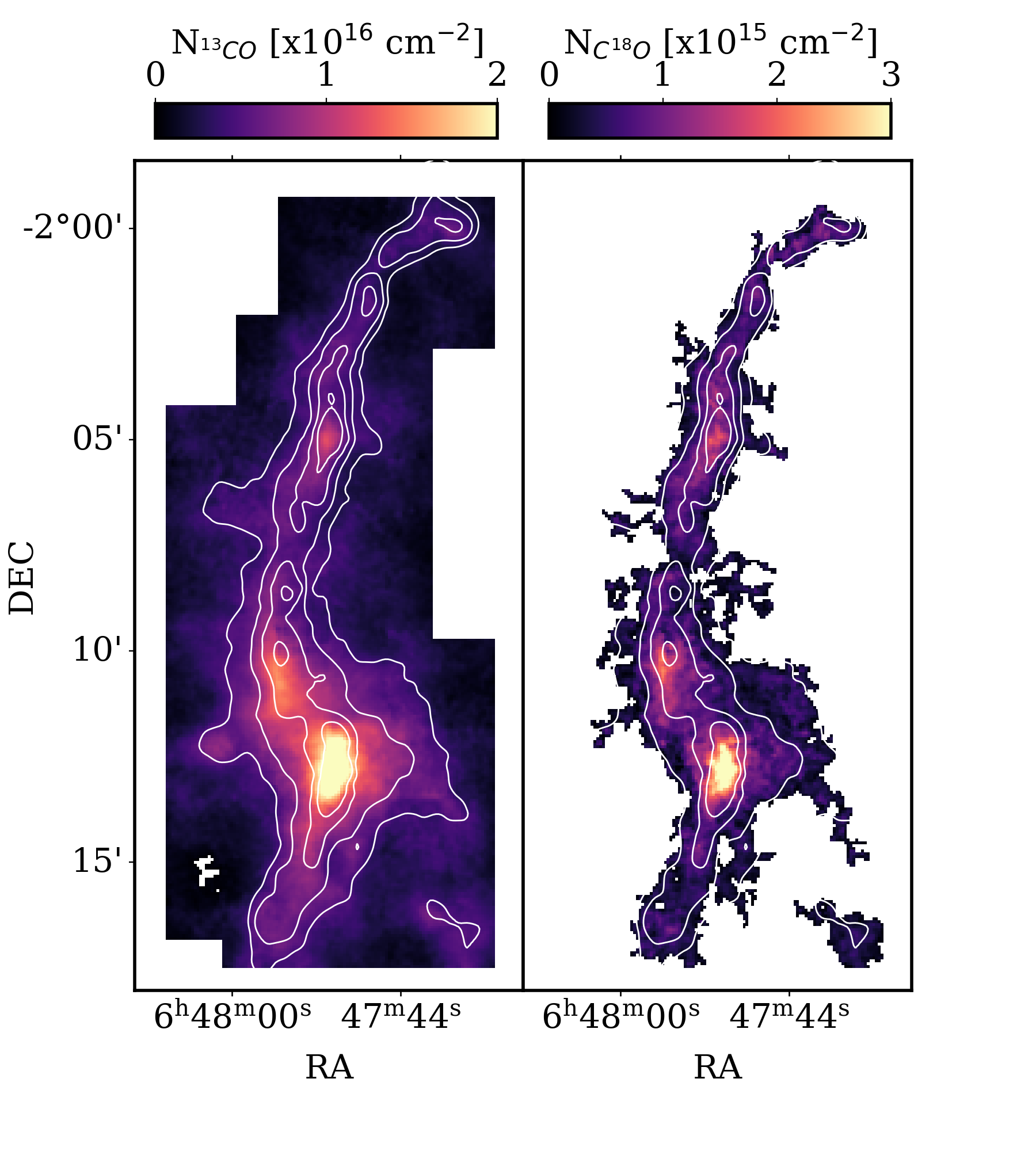}
\caption{A map of the column density of (left) $^{13}$CO and (right) C$^{18}$O. The contours denote total column density at an A$_v$ of 2, 3, 5 and 8 magnitudes.}
\label{fig::NCO}
\end{figure}  

We use the general formulation to estimate the column density from \citet{ManShi15} (their equation 84) which includes the effects of opacity:
\begin{multline}
N_{\rm tot} = \left( \frac{3h}{8\pi^3S\mu^2} \right) \left(\frac{Q(T_{\rm ex})}{g_u}\right) \left(\frac{e^{T_{ul}/T_{\rm ex}}}{ e^{T_{ul}/T_{\rm ex}} - 1}\right) \\ \times \int{-\ln{\left[1 - \frac{T_R}{J(T_{\rm ex}) - J(T_{bg})} \right]} dv},
\label{eq::Ncol}
\end{multline}
where $h$ is Planck's constant, $S$ is the line strength of the observed transition, $\mu$ is the dipole moment of the molecule, Q is the partition function, $g_u$ is the upper level degeneracy, and J is the function defined as $J(T) \equiv T_{ul} / (e^{T_{ul}/T} - 1)$. As we consider the (1-0) rotational transition, the level degeneracy is given as $g_u = 2J_u + 1$ and the line strength as $S = J_u / (2J_u + 1)$ where $J_u = 1$. Further, as $\coc$ and $\coo$ are diatomic linear molecules the rotational partition function may be approximated as
\begin{equation}
Q(T) \approx \frac{k_b T}{h B_0} + \frac{1}{3} + \frac{h B_0}{15k_b T},
\end{equation}
where $B_0$ is the rigid rotor rotation constant of the molecule and $k_b$ is Boltzmann's constant. The values we use for $B_0$ are 55101.01 MHz and 54891.42 MHz for $\coc$ and $\coo$ respectively \citep{Caz03, Caz04}. For the dipole moment we use 0.11046 Debye and 0.11049 Debye for $\coc$ and $\coo$ respectively \citep{Goo94}.

\begin{figure}
\centering
\includegraphics[width=0.8\linewidth, trim={0cm 3cm 0cm 3cm},clip]{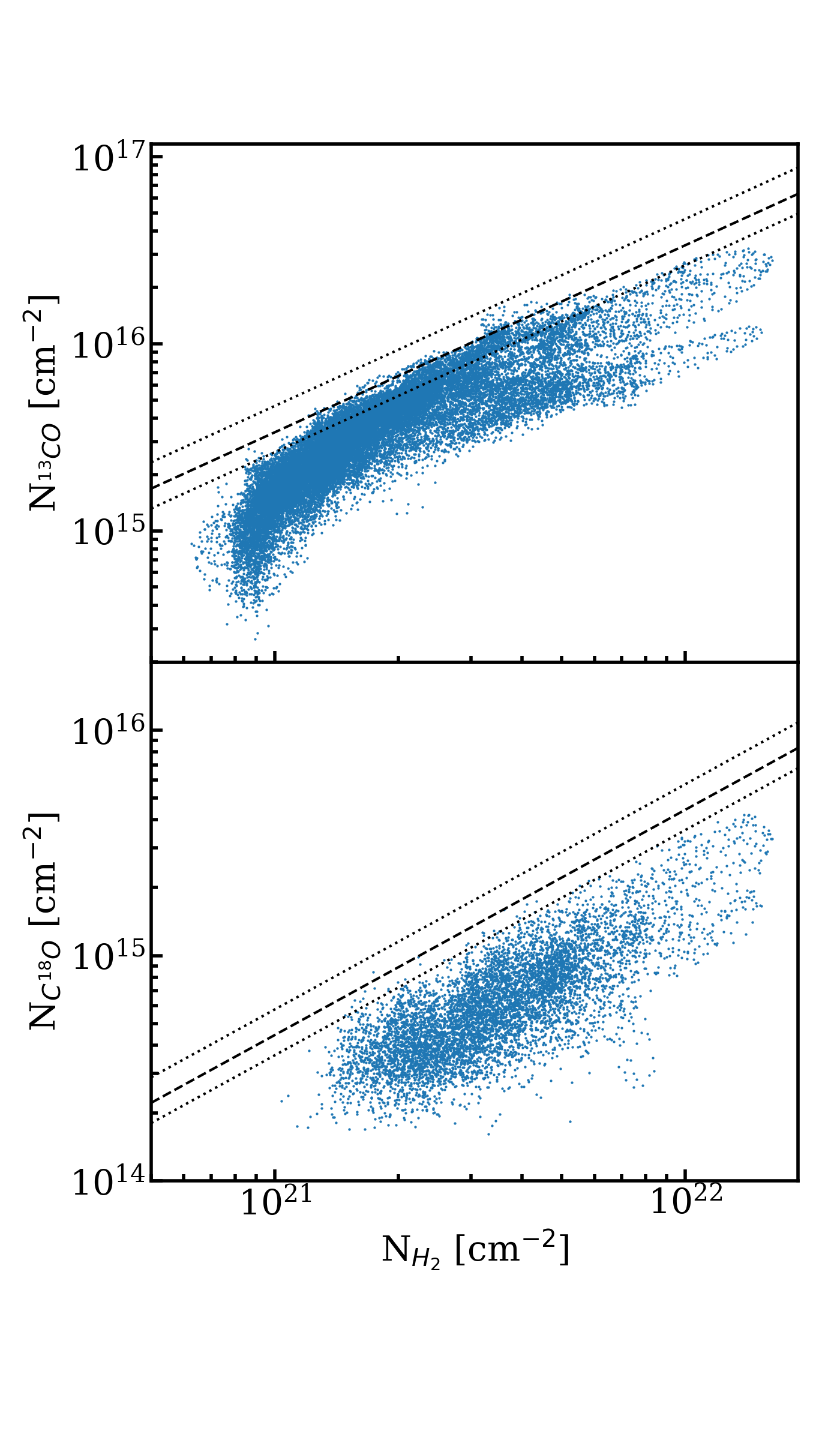}
\caption{Scatter plots showing the column density of (top) $^{13}$CO and (bottom) C$^{18}$O against the column density of H$_2$ as derived from \textit{Herschel}. Note the \textit{Herschel} column density values come from \citet{Cla23} and have had the unrelated line-of-sight column density subtracted. The dashed black lines show the expected abundances relative to H$_2$ using the \citet{WilRood94} isotope ratios at G214.5's Galactocentric distance of 10.1 kpc, i.e. $^{13}$CO/H$_2$ = 3.4$\times 10^{-6}$ and C$^{18}$O/H$_2$ = 4.4$\times 10^{-7}$. The dotted black lines show the 1-sigma uncertainties on the expected abundance.}
\label{fig::CD_NCO}
\end{figure}  

We evaluate equation \ref{eq::Ncol} using the same mask as that used to generate the moment maps to ensure that velocity channels containing only noise do not influence the calculation. The resulting column density maps for $\coc$ and $\coo$ are shown in figure \ref{fig::NCO}. One sees that the $\coc$ and $\coo$ column densities roughly follow the contours of the dust derived H$_2$ column densities, though there are some clear differences in the high column density clumps. The interquartile range of the error on the column density of $\coc$ is 3.9 - 8.4$\%$, while for $\coo$ it is 14.7 - 29.3$\%$ due to the much weaker emission. The opacities of these lines are also calculated; the median opacities of $\coc$ and $\coo$ are 0.34$^{+0.15}_{-0.10}$ and 0.07$^{+0.02}_{-0.02}$ respectively, showing that neither are highly optically thick. Maps of the $\coc$ and $\coo$ opacities, and their errors, are shown in the supplementary materials. One can see that even in the filament spine, the $\coc$ opacity only reaches values of $\sim$ 0.7. Assuming typical relative abundances between $\co$ and $\coc$ leads to high $\co$ opacities, $>12.5$ across 95$\%$ of the observed map and with a minimum value of $\sim5.2$, justifying the use of equation \ref{eq::tex} to calculate the excitation temperature even in the least dense regions.

To investigate further we show the $\coc$ and $\coo$ column density plotted against H$_2$ column density in figure \ref{fig::CD_NCO}. Note that here only pixels in which the CO column density has an error of less than 30$\%$ are shown. One sees a large increase in $\coc$ column density around N$_{\rm H_2} \sim 10^{21}$ cm$^{-2}$ (A$_v$ $\sim$ 1 mag), which may be attributed to the regime in which CO formation may efficiently proceed due to self-shielding \citep{Rol07,Bis19}. At higher column densities, the relationship is roughly linear indicating a constant abundance, though at even higher column densities ($> 3 \times 10^{21}$ cm$^{-2}$), two branches appear. For $\coo$ the relationship between $\coo$ column density and H$_2$ column density is simpler and appears to be roughly linear with scatter.

One may consider the abundance of $\coc$ and $\coo$ with respect to H$_2$ from these data. We use the isotopic gradient equations from \citet{WilRood94} to determine the expected abundance of $\coc$ and $\coo$. These are:
\begin{equation}
\frac{^{12}\rm{C}}{^{13}\rm{C}} = (7.5 \pm 1.9) R_{GC} + (7.6 \pm 12.9),
\label{eq::13C}
\end{equation}
and
\begin{equation}
\frac{^{16}\rm{O}}{^{18}\rm{O}} = (58.8 \pm 11.8) R_{GC} + (37.1 \pm 82.6),
\label{eq::18O}
\end{equation}
where $R_{GC}$ is the galactocentric distance in kiloparsec. At G214.5's galactocentric distance of 10.1 kpc this results in $^{12}\rm{C}/^{13}\rm{C} = 83.4 \pm 23.1$ and $^{16}\rm{O}/^{18}\rm{O} = 631.0 \pm 145.1$\footnote{\citet{Mil05} also provide an estimate for the $^{12}\rm{C}/^{13}\rm{C}$ ratio as a function of Galactocentric distance. For G214.5's distance this results in a ratio of 81.5 $\pm$ 12.5, consistent with the \citet{WilRood94} estimate.}. Assuming that the abundance of carbon relative to hydrogen is $1.4 \times 10^{-4}$ \citep{Ger15}, this leads to expected abundances of $\coc$ and $\coo$ with respect to H$_2$ of X$_{\coc} = 3.4^{+1.2}_{-0.7} \times 10^{-6}$ and X$_{\coo} = 4.4^{+1.4}_{-0.8} \times 10^{-7}$ \footnote{Here the $A^{+b}_{-c}$ notation does not describe the variance across all map pixels. Here it is used to show the propagation of the 1$\sigma$ uncertainty from equations \ref{eq::13C} and \ref{eq::18O} to the $\coc$ and $\coo$ abundances respectively due to their asymmetric nature.}. It is important to note that these expected abundances assumes no isotopic fractionation, all carbon is found in CO and that there is no depletion from the gas phase. Using these abundances, the expected relationship between the column density of $\coc$ and $\coo$, and H$_2$ is shown in figure \ref{fig::CD_NCO}. One sees that clearly, even considering the uncertainties, that the observed abundances of both isotopologues are mostly below the expected one. 

\begin{figure}
\centering
\includegraphics[width=0.8\linewidth, trim={1cm 3cm 0cm 0cm},clip]{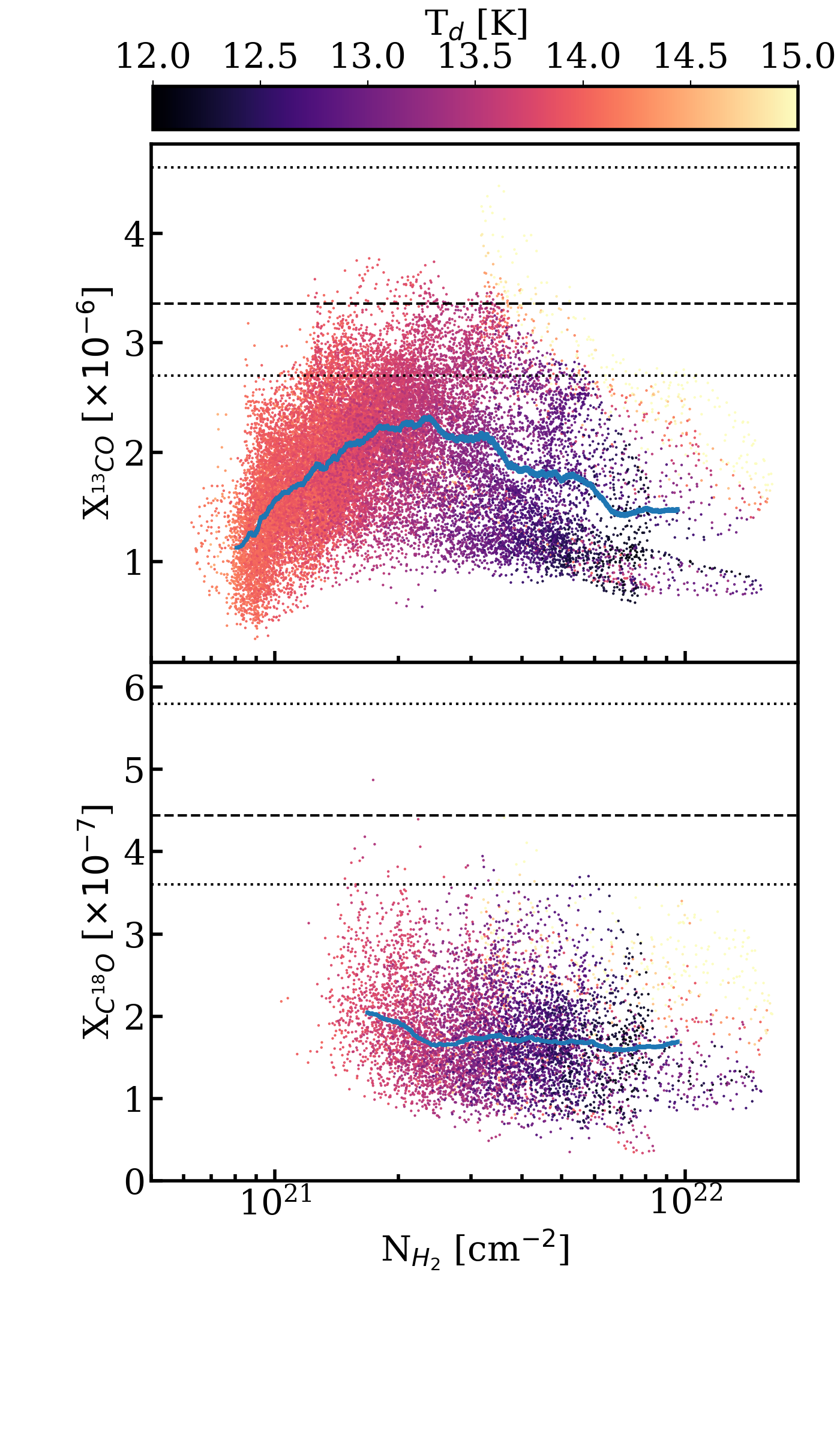}
\caption{Scatter plots showing the abundance of (top) $^{13}$CO and (bottom) C$^{18}$O with respect to H$_2$ against the column density of H$_2$ as derived from \textit{Herschel}, colour-coded by the \textit{Herschel} dust temperature. The dashed black lines show the expected abundances relative to H$_2$ using the \citet{WilRood94} isotope ratios at G214.5's galactocentric distance of 10.1 kpc, i.e. $^{13}$CO/H$_2$ = 3.4$\times 10^{-6}$ and C$^{18}$O/H$_2$ = 4.4$\times 10^{-7}$. The dotted black lines show the 1-sigma uncertainties on the expected abundance. The blue shaded regions show the error-weighted moving mean and its uncertainty. }
\label{fig::CD_XCO}
\end{figure}  

The abundance of $\coc$ in G214.5 is $1.9^{+0.4}_{-0.4} \times 10^{-6}$, which is approximately half what is expected. For $\coo$ the abundance is $1.7^{+0.5}_{-0.3} \times 10^{-7}$, nearly a third of what is expected. However, as seen in figure \ref{fig::CD_NCO}, the abundance is not independent of H$_2$ column density. As such we show X$_{\coc}$ and X$_{\coo}$ against H$_2$ column density in figure \ref{fig::CD_XCO}. For X$_{\coc}$, one sees the abundance increases until around A$_v$ $\sim 2$ mag, at which point it decreases with increasing H$_2$ column density until it is below $\sim 1 \times 10^{-6}$. For X$_{\coo}$, the abundance is roughly constant but there exists a slight decrease with increasing H$_2$ column density. This is confirmed by the use of Kendall's $\tau$ correlation test which reports $\tau = -0.12$ and $p \ll 10^{-10}$. As mentioned above, the increasing abundance of $\coc$ between an A$_v$ of 1 and 2 magnitudes indicates that this is the regime in which CO formation may efficiently proceed due to self-shielding. The decrease in abundance with increasing column density is most readily explained by CO depletion (or CO freeze-out), which is when gaseous CO molecules are locked onto the surfaces of dust grains in the form of ices \citep[e.g.][]{Cas02}. This is supported by the fact that at a given column density, the abundances for both $\coc$ and $\coo$ are lower with decreasing dust temperature. 

\begin{figure}
\centering
\includegraphics[width=0.92\linewidth, trim={0cm 1.5cm 2.2cm 0cm},clip]{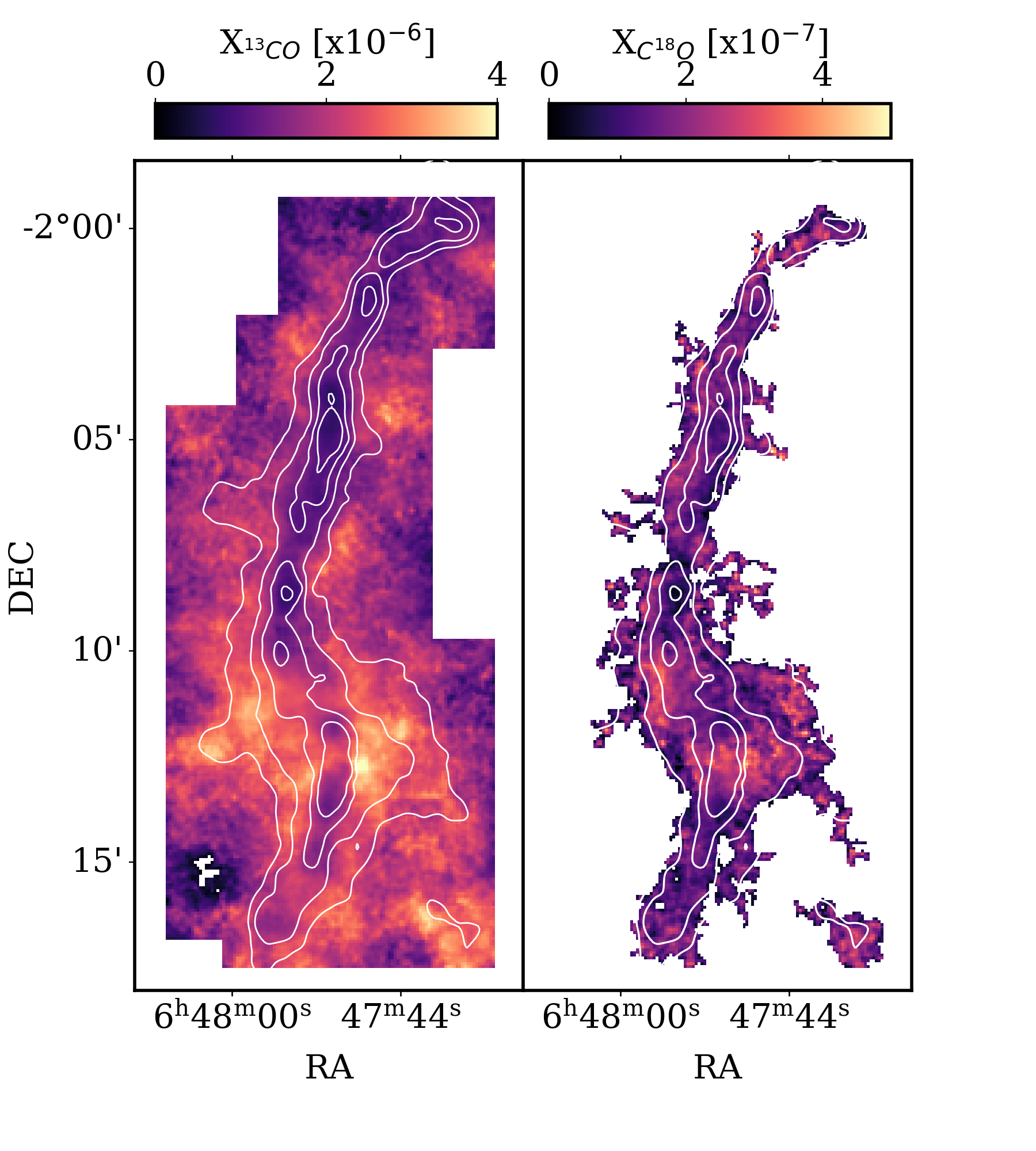}
\caption{A map of the abundance of (left) $^{13}$CO and (right) C$^{18}$O with respect to H$_2$. The contours denote total column density at an A$_v$ of 2, 3, 5 and 8 magnitudes.}
\label{fig::XCO}
\end{figure}  

To investigate the spatial distribution of X$_{\coc}$ and X$_{\coo}$, we show maps of these abundances in figure \ref{fig::XCO}. A striking feature of the $\coc$ map is that the low abundances ($\sim 1 \times 10^{-6}$) associated with freeze-out are not confined to small, localised regions but are seen along the entire $\sim 13$ pc long spine of G214.5. Typically such CO depletion is seen on the sub-parsec scale in cores and clumps \citep{Kra99,Ber02,Cas02,Bac03,Fon12,Sad15,Sab22}, though recent observations of infra-red dark clouds (IRDCs) have shown depletion on the parsec scale \citep{Her11,Sab19,Feng20}. Here, G214.5 shows CO depletion on the $>10$ pc scale across the whole cloud while not have column densities as high as typical IRDCs \citep{Cla23}. Thus such large scale depletion may be apparent in non-IRDC clouds as long as it is accompanied by widespread low gas and dust temperatures. This may be common in the Outer Galaxy and so ought to be considered. The X$_{\coo}$ map shows the abundance is roughly uniform across the filament, with possibly lower abundances in the northern section with its lower dust temperatures; however, the $\coo$ detection being limited to the spine of the filament makes it difficult to assess how depleted it is in the spine relative to material at larger radii.

\begin{figure}
\centering
\includegraphics[width=0.92\linewidth, trim={0cm 4cm 0cm 3.3cm},clip]{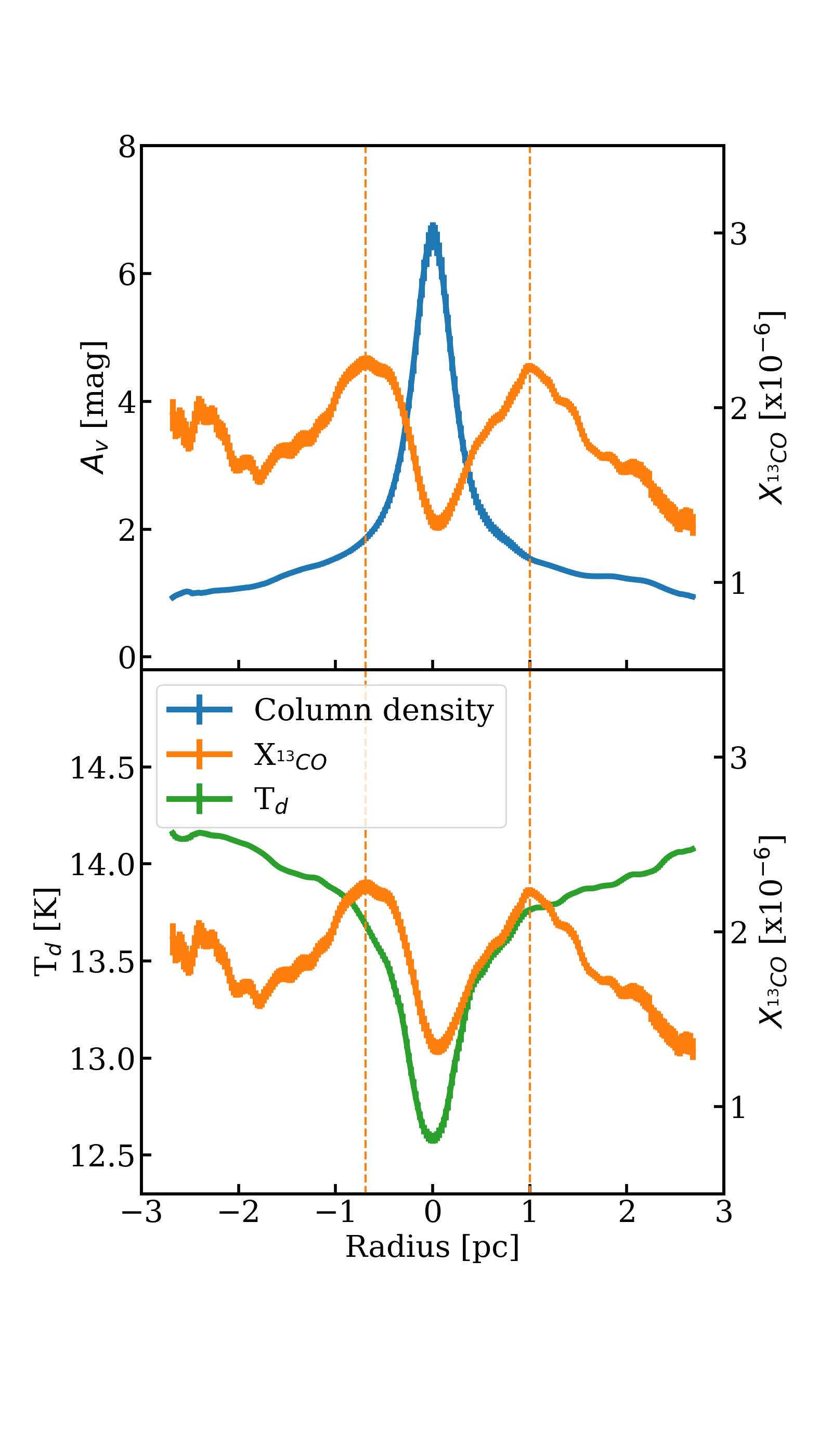}
\caption{(Top) The mean radial profile of (blue) column density, (orange) X$_{\coc}$, and (green) dust temperature. The shaded area denotes the error on the mean. The vertical, orange, dashed lines show the locations of peak $\coc$ abundance.}
\label{fig::XCO_rad}
\end{figure}

As the CO depletion zone is confined to the entire spine of G214.5, we measure the radial extent of the freeze-out by using the spine identified by \citet{Cla23} and projecting the abundance map into the orthogonal radial-longitudinal distance space with the \textsc{Python} module \textsc{FragMent} \citep{Cla19}\footnote{https://github.com/SeamusClarke/FragMent}. From this we construct the error-weighted mean abundance of $\coc$ as a function of radius, which is shown in figure \ref{fig::XCO_rad}, as well as the radial H$_2$ column density and dust temperature profiles. The weighting for the mean is taken as X$_{\coc}$/$\Delta$X$_{\coc}$, i.e. one over the uncertainty expressed as a fraction of the estimated abundance. Here one sees that the abundance profile is approximately flat at large radii ($>2$ pc) before it begins to rise slightly, peaking at $\sim$ 0.8-1.0 pc. Within this radius, it monotonically decreases, just as total column density begins to the increase rapidly and the dust temperature decreases strongly. Thus, CO freeze-out seems to occur out to $\sim$ 0.8 pc in G214.5. We note that the beamsize of the CO observations is 23.5'', or $\sim$ 0.26 pc at a distance of 2300 pc, is much smaller than this observed radius at which the abundance decreases. As such, the abundance profile cannot be due to freeze-out in only the inner, unresolved region but must be radially extended. 

\subsection{Modelling cloud-scale freeze-out}\label{SSEC:FREEZE}%

To further investigate the extent of this widespread freeze-out we construct an axisymmetric model of the H$_2$ and $\coc$ volume densities which may be constrained by the radial profiles of the H$_2$ column density and X$_{\coc}$. This model is constructed in 2D using the unprojected radius, $r$, before being projected to produce a column density as a function of projected radius, $R$. This allows the conversion between volume and column densities, taking into account that the degree of freeze-out differs along the line-of-sight, and correcting for beam convolution.

The H$_2$ volume density takes the form of a Plummer-like distribution, commonly used in characterising filaments \citep[e.g.][]{And14}, 
\begin{equation}
n_{H_2}(r) = \frac{n_c}{\left(1 + \left(\frac{r}{r_0} \right)^2 \right)^{p/2}} + n_{bg},
\label{eq::Plum}
\end{equation}
where $n_c$ is the central number density, $r_0$ is the inner, flattening radius, $p$ is the profile's exponent at large radii, and $n_{bg}$ is the background density. We calculate the number density out to a radius of 5 pc; this has little effect on the estimates of the model parameters except $n_{bg}$ which decreases as this outer radius increases. 

The volume density of $\coc$, which may be linked to the H$_2$ volume density by the equation
\begin{equation}
n_{\coc} = n_{H_2} X_{bg} f_{gas},
\end{equation}
where X$_{bg}$ is the abundance of $\coc$ with respect to H$_2$ unaffected by depletion, and $f_{gas}$ is the fraction of $\coc$ which is located in the gas phase. To calculate $f_{gas}$ we use the freeze-out model from \citet{GloCla16} which assumes an equilibrium between the rate at which CO accretes onto a dust grain, $R_{acc}$, the rate at which CO thermally evaporates from grains, $R_{therm}$, and the rate at which CO is desorbed due to cosmic rays, $R_{cr}$. This model produces the relation
\begin{equation}
f_{gas} = \frac{R_{therm} + R_{cr}}{R_{therm} + R_{cr} + R_{acc}}.
\end{equation}
These rates may be given as:
\begin{equation}
R_{acc} = 10^{-17} T_g^{0.5} n_H \left(\frac{D}{100}\right) \;\; \textrm{s}^{-1} \textrm{molecule}^{-1},
\end{equation}
where $T_g$ is the gas temperature, $n_H$ is the number density of hydrogen nuclei (i.e. $n_H = 2n_{H_2}$), and $D$ is the gas-to-dust ratio which we set to 100 here;
\begin{equation}
R_{therm} = 1.04 \times 10^{12} e^{-960/T_d} \;\; \textrm{s}^{-1} \textrm{molecule}^{-1},
\end{equation}
where $T_d$ is the dust temperature; and 
\begin{equation}
R_{cr} = 6 \times 10^{-13} \left(\frac{\zeta_H}{10^{-17}}\right) \;\; \textrm{s}^{-1} \textrm{molecule}^{-1},
\end{equation}
where $\zeta_H$ is the cosmic ray ionisation rate for atomic hydrogen. We assume that the rates of these reactions are identical for $\co$ and $\coc$.

Once the volume densities of H$_2$ and $\coc$ have been projected to column densities, $N_{\rm H_2}$ and $N_{\coc}$, these are convolved with a beam of size 23.5'' (i.e. the resolution of the observational data) assuming a distance of 2300 pc to G214.5. For the gas and dust temperatures we set $T_g$ to 8 K and $T_d$ to 13 K, informed by the gas temperatures found in section \ref{SSEC:TEX} and the dust temperatures of \citet{Cla23}. Our choices here have little effect on the final fitting; varying the dust temperatures in realistic ranges ($10$ K $\leq T_d \leq 15$ K) changes the fitting parameters by only a few percent, while changes to the gas temperature ($5$ K $\leq T_g \leq 10$ K) result in $\sim 10\%$ changes in only $\zeta_H$. Further, uniform temperatures may not be realistic (as seen by the radial dust temperature gradient seen in figure \ref{fig::XCO_rad}), however, using radial temperature profiles of the form $T(r) = T_0 + \Delta T \times r$ for both the gas and dust temperatures, also has minimal effect on the fitting parameters for a wide variety of gradients. This is because at these low dust temperatures, $R_{therm}$ is negligible and variations of $R_{acc}$ are dominated by changes in the number density, which covers orders of magnitudes, rather than the gas temperature which may vary by a factor of two at most.  With the temperatures constrained, this model has 6 parameters which we aim to constrain: $n_c$, $r_0$, $p$, $n_{bg}$, $X_{bg}$, $\zeta_H$.

\begin{figure}
\centering
\includegraphics[width=0.92\linewidth, trim={0cm 4cm 0cm 1.8cm},clip]{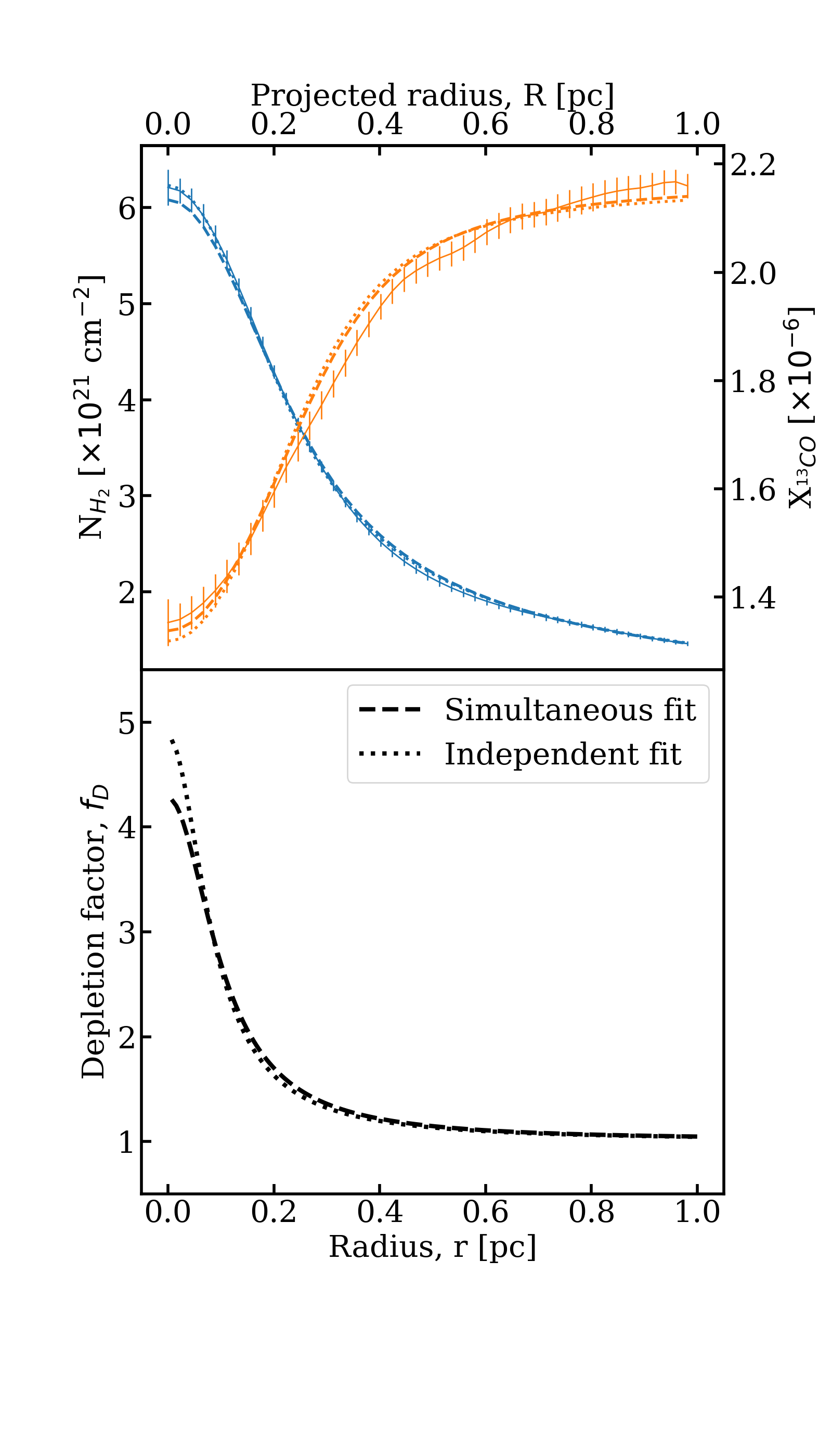}
\caption{(Top) The radial profiles of the (blue) column density and the (orange) $\coc$ abundance, X$_{\coc}$. The error bars show the error on the mean. The dashed lines show the results of the freeze-out model when simultaneously fitting both the column density and abundance profiles, and the dotted lines show the results when the two are independently fitted. (Bottom) The underlying radial profile of the depletion factor resulting from the model for the two fitting strategies.}
\label{fig::Dep_mod}
\end{figure}  

As mentioned above, the data used to constrain the model are the radial profiles of H$_2$ column density and the abundance of $\coc$, seen in figure \ref{fig::XCO_rad}. As the model is axisymmetric, we take the average of the left- and right-hand sides of the profiles to produce a single radial profile. We fit the column density profile out to a radius of 2 pc, to ensure that the exponent $p$ may be well-fitted and to avoid undue influence of further outside structures. The abundance profile is fitted out to a radius of 1 pc as it approximately peaks at this point; at larger radii the abundance drops and this cannot be modelled using our freeze-out model as this drop is likely due to CO formation rather than depletion. The model is fitted using a least-squared method and may be done in two manners: a simultaneous fitting of the column density and abundance profiles, and a independent fitting where the column density is fitted first before using the result to fit the abundance profile. We present the results of these fits in figure \ref{fig::Dep_mod} as well as the best-fitting parameters in table \ref{tab::Dep_mod}.

\begin{table}
\centering
\begin{tabular}{@{}*3l@{}}
\hline\hline
  &  Simultaneous  & Independent  \\ \hline
$n_c$ [cm$^{-3}$] & 6886 $\pm$ 295 & 8707 $\pm$ 354 \\
$r_0$ [pc] & 0.103 $\pm$ 0.004 & 0.083 $\pm$ 0.003 \\
$p$ & 2.00 $\pm$ 0.03 & 1.91 $\pm$ 0.02 \\
$n_{bg}$ [cm$^{-3}$] & 26.5 $\pm$ 0.4 & 25.3 $\pm$ 0.3 \\
$X_{bg}$ [$\times 10^{-6}$] & 2.199 $\pm$ 0.006 & 2.188 $\pm$ 0.009 \\
$\zeta_H$ [$\times 10^{-18}$ s$^{-1}$] & 1.860 $\pm$ 0.006 & 1.996 $\pm$ 0.008 \\\hline \hline
\end{tabular}
\centering
\caption{A table showing the fitting parameters of the axisymmetric freeze-out model when fitting the radial column density and $\coc$ abundance profiles simultaneously or independent of each other. Note that the uncertainties quoted here are from the fitting alone; uncertainties due to dust temperature and gas temperature will lead to $\sim$10$\%$ uncertainties on $X_{bg}$ and $\zeta_H$.}
\label{tab::Dep_mod}
\end{table}

One sees from figure \ref{fig::Dep_mod} that this simple axisymmetric freeze-out model well produces the observed column density and abundance profiles. Both methods of fitting produce relatively similar profiles and fitting parameters, with independent fitting leading to a slightly larger central volume density and smaller flattening radius. The independent fitting method leads to a moderately better fit to the column density profile, $\chi^2$ = 25.0 compared to 42.7 for the simultaneous fitting method, while the fit to the abundance profile is moderately worse, $\chi^2$ = 95.5 compared to 52.9. 

An interesting result from the model is that due to the relatively low central densities determined, a low cosmic-ray ionisation rate is needed to ensure the deep and wide depletion feature seen in the $\coc$ abundance profile, with best-fitting values of $\sim 1.9 \times 10^{-18}$ s$^{-1}$. This is approximately an order of magnitude lower than the rate which is typically observed, a few times $10^{-17} - 10^{-16}$ s$^{-1}$ \citep{deB96,Cas98,Van00,IndMcc12,Ind15}, although there have been some measurements of the order of a few times $10^{-18}$ s$^{-1}$ in a selection of dense cores \citep[e.g.][]{Cas98}. Further, this low cosmic-ray ionisation rate is consistent with the low gas temperatures we find in section \ref{SSEC:TEX}. 

Cosmic-ray ionisation rates of the order of $2-5 \times 10^{-18}$ s$^{-1}$ have also been estimated in the IRDC G28.37+00.07 by \citet{Ent22} using a gas-grain astrochemical model towards 10 lines-of-sight constrained by numerous molecular line observations. This is a highly different environment to G214.5 as the IRDC is much denser ($n>5-10\times10^{4}$ cm$^{-3}$) and at higher column densities ($N>10^{23}$ cm$^{-2}$). It is possible that the low cosmic ray ionisation rate observed by \citet{Ent22} may be the result of the attenuation of the lower energy cosmic-rays (i.e. those important for ionisation and chemistry) due to the high column densities in G28.37+00.07. Such attenuation of these lower energy cosmic-rays is predicted by cosmic-ray transport models \citep[e.g.][]{Pad09,Sch16,Ivl18,Pad18,Gac22}. However, attenuation is unlikely the reason for the measured low cosmic-ray ionisation rate in G214.5 as the column densities at which freeze-out occurs is nearly two orders of magnitude lower ($N\sim2-3\times10^{21}$ cm$^{-2}$). Thus, the measured cosmic-ray ionisation rate in G214.5 is likely close to the value in G214.5's diffuse environment.

As the model constructs an underlying radial profile, unaffected by projection or beam convolution, we may investigate the radial profile of the depletion factor, $f_D = 1/f_{gas}$. This is shown in the lower panel of figure \ref{fig::Dep_mod}. Both fitting procedures lead to strongly centrally condensed depletion factor profiles, with peak values $\sim5$, i.e. only $\sim$20$\%$ of the $\coc$ is in the gas phase. The radius at which half of the $\coc$ is in the gas phase, $f_D=2$, is $\sim 0.16$ pc in both models, and the radius at which two thirds is in the gas phase, $f_D=1.5$ is $\sim 0.30$ pc. We may also determine the number density at these radii, $\sim 2200$ cm$^{-3}$ and $\sim 720$ cm$^{-3}$ respectively. Thus, tracing even moderately dense gas using CO is affected by depletion. Using these volume densities, one may determine an estimate of the timescale over which this depletion occurs, $\tau_{D} = 1/R_{acc} \sim 560 \;\rm{Myr} / n_{H_2}$. The central density of $\sim 8000$ cm$^{-3}$ yields a depletion timescale $\sim 0.07$ Myr, and for the $f_D=2$ density of $\sim 2200$ cm$^{-3}$, $\tau_D \sim 0.25$ Myr. These are much lower than the cloud free-fall time and cloud-crushing timescales calculated by \citet{Cla23} for G214.5, 13 Myr and 2-3 Myr respectively, showing that such depletion occurs sufficiently quickly for it to be feasible and consistent with the expected youth of G214.5.

Comparing to \citet{Her11} and \citet{Sab19}, which study the parsec-scale CO depletion in the filamentary IRDCs G035.39-00.33 and G351.77-00.51 respectively, we find similar peak depletion factors, $f_D \sim 5$. Further, \citet{Sab19} also model the radial extent of the CO depletion using multiple imposed abundance profiles, and find a depletion radius in the range 0.02-0.15 pc, slightly smaller but comparable to the $f_D=2$ radius we see in G214.5. Thus, the large-scale CO depletion along the spines of cold, giant filamentary clouds may be commonplace. However, here we find that depletion begins at volume densities an order of magnitude lower than \citet{Her11} and \citet{Sab19}, who find depletion to become important above $n_{\rm H_2} \sim 2-50 \times 10^{4}$ cm$^{-3}$. Further, freeze-out starts to appear in G214.5 at column densities of only $\geq 2 \times 10^{21}$ cm$^{-2}$ compared to $\geq 1 \times 10^{22}$ cm$^{-2}$ in the aforementioned IRDCs. This shift reflects the very low temperatures in G214.5, even off of the spine, related to its quiescent environment characterised by a low radiation field and cosmic-ray ionisation rate. This may suggest that in other quiescent environments (e.g. in the Outer Galaxy) one might expect to see widespread CO depletion at lower volume and column densities. A consequence of this is that estimates of masses and column densities made without considering cloud-scale CO freeze-out will be underestimates; for the case of G214.5, if one used the expected abundance of $\coc$, X$_{\coc} = 3.4 \times 10^{-6}$, one would underestimate the total mass by roughly a factor of 2, and the mass at high column densities would be reduced by roughly a factor 3 due to decreasing X$_{\coc}$ with increasing column density (figure \ref{fig::CD_XCO}). Further, considering the kinematics of CO isotopologues, even moderately dense gas might not be probed as expected and measurements of velocity centroids/dispersion affected. 

\subsubsection{Ambipolar diffusion}\label{SSSEC:AMBI}%

As well as the low cosmic-ray ionisation rate allowing widespread CO depletion to occur, it will have the consequence of lowering the ionisation fraction of the gas as cosmic-rays are the dominant ionisation source in UV-shielded gas \citep{Wol22}. Low ionisation fractions have the consequence of imperfect coupling between the gas and the magnetic field, meaning that non-ideal MHD effects like ambipolar diffusion become important \citep[e.g.][]{Nto16,Zhao18,Wur19,Wur20,Wur21,Zhao21,Com22}. 

To consider if this may be relevant for G214.5, we use the \textsc{Nicil} code presented in \citet{Wur16} to self-consistently determine the ionisation fraction and ambipolar diffusivity, $\eta_{\rm AD}$, at a given density, temperature, cosmic-ray ionisation rate and magnetic field strength. For the density we use the average density within the full-width half-maximum (FWHM) of the filament model profile described above, FWHM = 0.17-0.21 pc and $\langle n \rangle$ = 5500-6900 cm$^{-3}$, where the variation comes from the two fitting methods. We use a range of gas temperatures of 5-10 K and a cosmic-ray ionisation rate of $1.8-2.0 \times 10^{-18}$ s$^{-1}$. Further, we use \textsc{Nicil} with a grain-size distribution proportional to $a^{-3.5}$ between a range of 0.03 $\mu$m - 0.1 cm, discretised into 100 equal size bins. The B-field strength is unknown for G214.5 and not constrained; we thus use a range between 20-200 $\mu$G which are typical field strengths found across IRDCs \citep{Hac23}\footnote{Taken from their table 2 where they report an interquartile range of 45-500 $\mu$G for IRDCs which have a line-mass interquartile range of 31-115 M$\Su$/pc.}. One may construct an ambipolar timescale from these diffusivities which is related to the timescale over which magnetic flux is lost via ion-neutral drift,
\begin{equation}
\tau_{\rm AD} = \frac{L^2}{\eta_{\rm AD}},
\end{equation}
where L is the length-scale considered \citep{Pat23}; here we use the FWHM of the filament model, 0.17-0.21 pc. The resulting ambipolar diffusivities and timescales are summarised in table \ref{tab::etaad}. One sees that the timescale ranges from $\sim0.35-45$ Myr, with shorter timescales corresponding to higher magnetic field strengths. Increasing the cosmic-ray ionisation rate by an order of magnitude ($1.9 \times 10^{-17}$ s$^{-1}$), i.e. within the typical range in other regions, the ambipolar timescale is increased by a factor of 4-5.

To compare to the ambipolar diffusion timescale, we may consider the cloud free-fall time and cloud crushing time calculated by \citet{Cla23}, 13 Myr and 2-3 Myr respectively, for estimates of cloud-scale timescales. For smaller scales, we may consider the thermal and turbulent crossing timescales, $L/c_s$ and $L/\sigma_{NT}$, where the sound speed $c_s=0.16$ km/s for a gas temperature of 8.2 K and $\sigma_{NT} = 0.35$ km/s taken from typical line-widths of $\coo$ along the filament spine using the 0.2 km/s resolution VESPA data. This results in timescales of 1.1-1.3 Myr and 0.5-0.6 Myr respectively. Taken together, the ambipolar diffusion timescale is shorter than/comparable to cloud-scale timescales for magnetic field strengths above 50 $\mu$G and is shorter than/comparable to the crossing times when above 100 $\mu$G. Thus, even a low-to-moderate magnetic field strength results in ambipolar diffusion being an important process in the spine of G214.5 due to the low cosmic-ray ionisation rate.  

\begin{table}
\centering
\begin{tabular}{llr}
\hline\hline
B-field [$\mu$G] &  $\eta_{\rm AD}$ [cm$^2$ s$^{-1}$] & $\tau_{\rm AD}$ [Myr]  \\ \hline
20 & $2.2-3.9 \times 10^{20}$  & 29.4-45.0 \\
50 & $1.4-2.4 \times 10^{21}$ & 4.7-7.2 \\
100 & $5.5-9.8 \times 10^{21}$ & 1.2-1.8 \\
200 & $2.2-3.9 \times 10^{22}$ & 0.3-0.4 \\\hline \hline
\end{tabular}
\centering
\caption{A table showing the ambipolar diffusivity, $\eta_{\rm AD}$, calculated from $\textsc{Nicil}$ and the corresponding ambipolar timescale, $\tau_{\rm AD}$, for 4 different B-field strengths.}
\label{tab::etaad}
\end{table}

\subsection{The isotopologue ratio between $\coc$ and $\coo$}\label{SSEC:X1318}%
\begin{figure}
\centering
\includegraphics[width=0.92\linewidth, trim={0cm 1.5cm 2.2cm 0cm},clip]{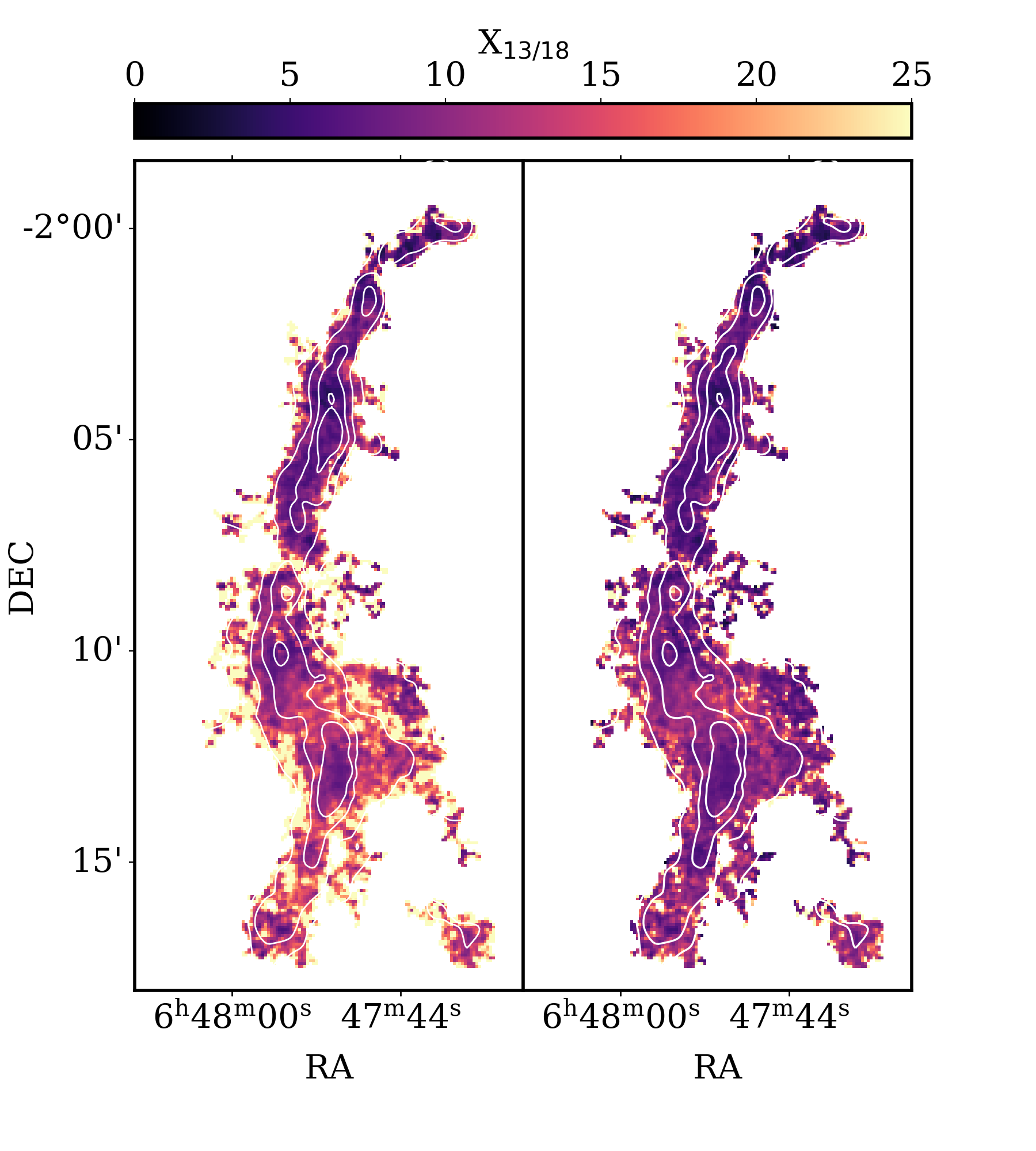}
\caption{A map of the abundance ratio of $^{13}$CO over C$^{18}$O, X$_{13/18}$, when using (left) the total $^{13}$CO column density and (right) the $^{13}$CO column density in velocity channels with C$^{18}$O emission. The contours denote total column density at an A$_v$ of 2, 3, 5 and 8 magnitudes.}
\label{fig::X1318}
\end{figure}  

An advantage of the method used in section \ref{SSEC:NCO} to determine the column densities of $\coc$ and $\coo$ is that the two are independently calculated, allowing a study of the isotopologue ratio X$_{13/18}$ = N$_{\coc}$/N$_{\coo}$. This isotopologue ratio is known to vary between clouds and within clouds for a number of reasons, such as local isotope enrichment, selective photo-dissociation and selective fractionation, and ranges from $\sim$4-50 \citep[e.g.][]{Are18,Are19,Lin21,Rou21}. One reason for the variation between clouds is the dependence of the $^{12}$C/$^{13}$C and $^{16}$O/$^{18}$O isotope ratios on Galactocentric radius mentioned above (equations \ref{eq::13C} and \ref{eq::18O}). Combining these two equations results in
\begin{equation}
\frac{^{13}\rm{C}^{16}\rm{O}}{^{12}\rm{C}^{18}\rm{O}} = \frac{(58.8 \pm 11.8) R_{GC} + (37.1 \pm 82.6)}{(7.5 \pm 1.9) R_{GC} + (7.6 \pm 12.9)},
\label{eq::X1318}
\end{equation}
which for G214.5's Galactocentric radius of 10.1 kpc is equal to $7.6^{+2.1}_{-1.6}$ $\;$ \footnote{Here the $A^{+b}_{-c}$ notation does not describe the variance across all map pixels. Here it is used to show the propagation of the 1$\sigma$ uncertainty from equations \ref{eq::X1318} due their asymmetric nature.}. Using the column densities of $\coc$ and $\coo$ calculated above produce a map of the isotopologue ratio X$_{13/18}$ shown in the left panel of figure \ref{fig::X1318}. We find a median value of $13.5^{+7.6}_{-3.9}$ and a median uncertainty of 25$\%$; these observed ratios are larger than the expected isotopologue ratio given G214.5's Galactocentric radius. Further, as seen in figure \ref{fig::X1318} there are spatial variations of the ratio with higher values towards the edge of the filament spine, i.e. at lower column densities.

When calculating the above isotopologue ratio the whole detected line-of-sight $\coc$ and $\coo$column densities are considered. However, due to the lower brightness of the $\coo$ emission and the noise level of the observations, the detected $\coo$ column density is likely tracing a less extended region of gas along the line-of-sight compared to the detected $\coc$ column density. Closer inspection of the spectra reveal that multiple velocity components are present in $\coc$ (predominately in the south of the filament around the massive clump C11), which are not detected in the $\coo$ spectra. Thus, using the total detected line-of-sight column densities to determine the isotopologue ratio will bias the result to larger values. To correct for this bias we recalculate the $\coc$ column density using the mask derived from the $\coo$ data cube, i.e. only using velocity channels in which $\coo$ is detected. This attempts to ensure that the two column densities used to calculate the isotopologue ratio are tracing the same line-of-sight gas. The resulting ratio is shown in the right panel of figure \ref{fig::X1318}, and yields a median ratio of $9.4^{+2.9}_{-2.1}$. This value is only slightly elevated compared to the expected ratio of $7.6^{+2.1}_{-1.6}$ but consistent given the uncertainties. When considering only pixels which have uncertainties less than 30$\%$ the median corrected ratio lowers further to $8.2^{+1.8}_{-1.4}$. Moreover, as shown in figure \ref{fig::X1318} the spatial variation in the ratio is much reduced. This shows the importance of using the kinematic information to correct for this bias and ensure that the $\coc$ and $\coo$ column densities used to calculate the isotopologue ratio are tracing the same gas.

\begin{figure}
\centering
\includegraphics[width=0.8\linewidth, trim={0.7cm 0cm 2cm 1.3cm},clip]{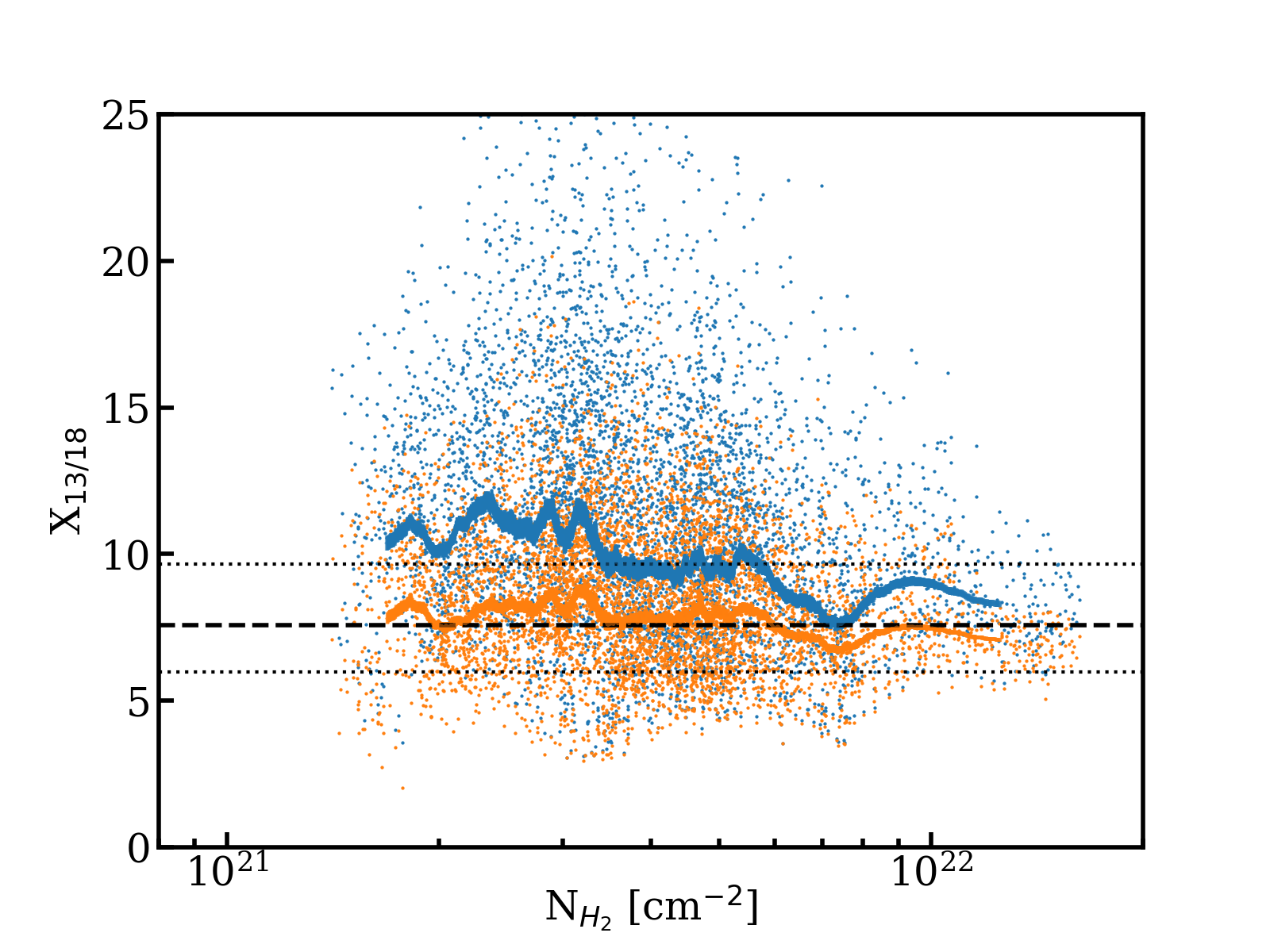}
\caption{Scatter plot showing the abundance ratio of $^{13}$CO over C$^{18}$O, X$_{13/18}$, against the column density of H$_2$ as derived from \textit{Herschel}, when using (blue) the total $^{13}$CO column density and (orange) the $^{13}$CO column density corresponding to C$^{18}$O emission. Only pixels where the error is below 30\% are used. The shaded regions shows the weighted moving mean with its corresponding uncertainty for the distribution of the same colour. The black, dashed line shows the X$_{13/18}$ value of 7.6 expected from \citet{WilRood94} at G214.5's Galactocentric distance of 10.1 kpc, with the dotted lines show the interquartile range considering the uncertainties in equation \ref{eq::X1318}.}
\label{fig::CD_X1318}
\end{figure}

To further investigate the variation across the cloud we show the isotopologue ratio, both the total column density ratio and the corrected ratio, as a function of column density in figure \ref{fig::CD_X1318}. One sees a large spread in the ratio (mostly due to uncertainties in the ratio which are typically $\sim1-2$), especially in the total column density ratio, making a search for a correlation difficult. A better view is seen when using the error-weighted moving-mean (the solid lines) which reveals that the corrected ratio is lower on average (by $\sim2-4$) for all column densities and lies extremely close to the expected ratio independent of column density. Typically an anti-correlation is seen between the isotopologue ratio and column density due to the selective enhancement of $\coc$ due to fractionation and destruction of $\coo$ via photodissociation at low column densities \citep[e.g.][]{Lan80,Are19,Rou21,Rod21}. However, the lack of such a correlation in G214.5 suggests that even the poorly shielded gas is relatively unaffected by the interstellar UV field, indicating that the background radiation field in G214.5's vicinity is low. This is in agreement with the low upper limit placed on the ambient FUV-field by the non-detection of [CII] (section \ref{SSEC:CII}), as well as the low dust/gas temperatures, the presence of $\co$ and $\coc$ at low $A_v$, as well as the fact there are no nearby OB stars/cluster \citep{Lee94,Wri23}.

\section{Conclusions}\label{SEC:CON}%

In this paper we present CO observations taken from the IRAM 30m telescope of the northern region of the giant molecular filament G214.5-1.8. We use these data to study the CO excitation temperature, the abundances of $\coc$ and $\coo$, as well as isotopologue ratio X$_{13/18}$, revealing G214.5 to be a highly interesting object. 

We use the $\co$ (1-0) and (2-1) lines to estimate the excitation temperature and find a relatively uniform and low value across the entire mapped region, $8.2^{+0.7}_{-0.6}$ K, with moderately elevated excitation temperatures close to the two observed outflows (section \ref{SSEC:TEX}). The excitation temperatures of these two lines are near identical, showing that the low excitation temperatures are not due to sub-thermal excitation but rather the gas is thermalised and as such the excitation temperatures are representative of low gas temperatures ($\sim$7-9 K). This is in agreement with the low dust temperatures in G214.5 found by \citet{Cla23} who find the cloud to be the coldest GMF yet studied. Moreover, widespread low gas/dust temperatures at low column densities are in agreement with the upper limit placed on the ambient FUV-field by the non-detection of [CII], $\lesssim 1.1-2.5$ G$_0$ (section \ref{SSEC:CII}). It is potentially due to G214.5's position in the Outer Galaxy that such temperatures and heating rates are found. As such further studies of the thermal properties of GMFs in the Outer Galaxy and their comparison to those in the Inner Galaxy would be valuable.

Using the derived excitation temperatures we calculate the column densities of $\coc$ and $\coo$ independent of each other, i.e. no assumption is made about isotopologue ratios (section \ref{SSEC:NCO}). Further, we use the column density map produced from \textit{Herschel} by \citet{Cla23} to calculate abundances for $\coc$ and $\coo$ with respect to H$_2$, X$_{\coc}$ and X$_{\coo}$. We find abundances of X$_{\coc}$=$1.9^{+0.4}_{-0.4} \times 10^{-6}$ and X$_{\coo}$ = $1.7^{+0.5}_{-0.3} \times 10^{-7}$; both are slightly lower than but similar to that found in nearby clouds \citep[e.g.][]{Pin10,Rou21}. Investigating the $\coc$ abundance further, we find that it increases with column density below an $A_v \sim 2$ mag, possibly indicating formation below this visual extinction.

We find that above an $A_v \sim 2$ mag, the $\coc$ abundances drops with increasing column density and this decrease coincides with a decreasing dust temperature. Due to this we suggest that this is evidence of CO freeze-out in cold, dense gas. Investigating the spatial extent of this drop in the $\coc$ abundance we see that it is found along the entire $\sim 13$ pc filament spine and out to a radius of roughly 0.8 pc, making G214.5 a rare case of cloud-scale CO freeze-out.

To further investigate the widespread freeze-out we construct a CO depletion model considering CO accretion, thermal desorption and cosmic-ray induced desorption. In concert with a model of the underlying volume density, we use this depletion model to fit the mean radial profiles of G214.5's column density and $\coc$ abundance (section \ref{SSEC:FREEZE}). We find that, to reproduce the large radial extent of the depletion, a very low cosmic-ray ionisation rate is required, $1.8-2.0 \times 10^{-18}$ s$^{-1}$. This is due to freeze-out occurring at low volume densities ($f_{gas}$=0.5 at r$\sim0.2$ pc, n$\sim2000$ cm$^{-3}$), which is unique to G214.5. An inferred cosmic-ray ionisation rate of the order of 10$^{-18}$ s$^{-1}$ is an order of magnitude lower than is typically measured, but is consistent with the very low gas and dust temperatures seen in G214.5. It is possible that other clouds in the Outer Galaxy may experience such low cosmic-ray ionisation rates due to their Galactocentric radii and quiescent environments \citep[e.g.][find a decrease in cosmic-ray flux with increasing galactocentric radius using gamma-ray observations]{Yang16}, making cloud-scale CO freeze-out more common. Such cloud-scale CO freeze-out has consequences on mass and column density estimates, as well as the interpretation of kinematic signatures, and it is therefore imperative to better understand the prevalence of this phenomenon. 

An additional consequence of the low cosmic-ray ionisation rate found in G214.5 is that it leads to a low ionisation fraction of the gas. This results in imperfect coupling between the predominately neutral gas and the magnetic field and the growing importance of non-ideal MHD terms. To study this we use the \textsc{Nicil} code \citep{Wur16} to self-consistently determine the ambipolar diffusivity given the results of our depletion model (section \ref{SSSEC:AMBI}). From this we estimate an ambipolar diffusion timescale for the filament spine and find that even a low-to-moderate magnetic field strength of above 50-100 $\mu$G is sufficient to result in an ambipolar diffusion timescale comparable to or smaller than other relevant timescales such as the cloud free-fall time, cloud crushing time as well as the sound and turbulent crossing time. Thus, it is possible that ambipolar diffusion is an important physical process on the relatively large-scale of G214.5's 13 pc spine. 

As the column densities of $\coc$ and $\coo$ are calculated independently of each other, we determine the isotopologue ratio between them, X$_{13/18}$ (section \ref{SSEC:X1318}). We find a large spread in the measured ratio, X$_{13/18}$ = $13.5^{+7.6}_{-3.9}$, and that it is elevated compared to the expected ratio considering G214.5's Galactocentric radius, $7.6^{+2.1}_{-1.6}$. However, due to the lower brightness of the $\coo$ emission and presence of noise, the detected $\coo$ column density is likely tracing a less extended region of gas along the line-of-sight compared to the detected $\coc$ column density, and thus introducing a bias elevating the observed abundance ratio. To address this bias, we calculate a corrected ratio that only considers velocity channels in which both are detected; this results in a lower ratio with considerably less spread which is consistent with the expected ratio, X$_{13/18}$ = $9.4^{+2.9}_{-2.1}$. It is therefore important to use the kinematic information to ensure that the measured isotopologue ratio is not biased by the non-detection of $\coo$ in some of the velocity channels in which $\coc$ is detected. Further, we show that X$_{13/18}$ is relatively independent of the column density, indicating that fractionation and selective photodissociation are only having minor effects on CO isotopologue abundances, suggesting a low interstellar radiation field in the vicinity of G214.5, consistent with the low upper limit placed using the non-detection of [CII].

Taken together, our results show that G214.5 is a highly interesting object of study which may be used to understand the star formation process in a quiescent region. Low gas and dust temperatures as well as large-scale CO freeze-out may be common in such quiescent regions and should be considered, as well as their effects on CO observations. Further studies of G214.5 to investigate the physical and chemical consequences of its low cosmic-ray ionisation rate and corresponding cloud-scale CO freeze-out will be the focus of future works. 

\section{Acknowledgments}\label{SEC:ACK}%
The authors would like to thank the anonymous referee for their helpful and constructive comments. SDC would also like to thank James Wurster for kindly answering a number of questions about ambipolar diffusion and the \textsc{Nicil} code. SDC is supported by the Ministry of Science and Technology (MoST) in Taiwan through grant MoST 108-2112-M-001-004-MY2. A.S.M. acknowledges support from the RyC2021-032892-I grant funded by MCIN/AEI/10.13039/501100011033 and by the European Union ’Next GenerationEU’/PRTR, as well as the program Unidad de Excelencia María de Maeztu CEX2020-001058-M and the grant PID2020-117710GBI00 (MCI-AEI-FEDER, UE). This work includes observations carried out under project number 009-19 with the IRAM 30m telescope. IRAM is supported by INSU/CNRS (France), MPG (Germany) and IGN (Spain). This publication includes data acquired with the Atacama Pathfinder Experiment (APEX) under programme ID O-0103.F-9318A-2019. APEX is a collaboration among the Max-Planck-Institut fur Radioastronomie, the European Southern Observatory, and the Onsala Space Observatory. This work includes observations made with the NASA/DLR Stratospheric Observatory for Infrared Astronomy (SOFIA) under project ID 83\_0723. SOFIA is jointly operated by the Universities Space Research Association, Inc. (USRA), under NASA contract NAS2-97001, and the Deutsches SOFIA Institut (DSI) under DLR contract 50 OK 0901 to the University of Stuttgart. SDC would also like to acknowledge the people developing and maintaining the open source packages which were used in this work: \textsc{Matplotlib} \citep{matplotlib}, \textsc{NumPy} \citep{numpy}, \textsc{SciPy} \citep{scipy} and \textsc{AstroPy} \citep{astropy}.

\section{Data availability}\label{SEC:DATA}%
The data underlying this article will be shared on reasonable request to the corresponding author. The analysis tools of this work, \textsc{BTS} and \textsc{FragMent}, are made freely available at https://github.com/SeamusClarke/FragMent and https://github.com/SeamusClarke/BTS.

\bibliographystyle{mn2e}
\bibliography{ref} 

\begin{appendices}

\section{Constraining the FUV field using SOFIA [CII] observations}\label{APP:CII}%
\begin{figure*}
\centering
\includegraphics[width=0.8\linewidth, trim={0.75cm 0.45cm 3.1cm 0.4cm},clip]{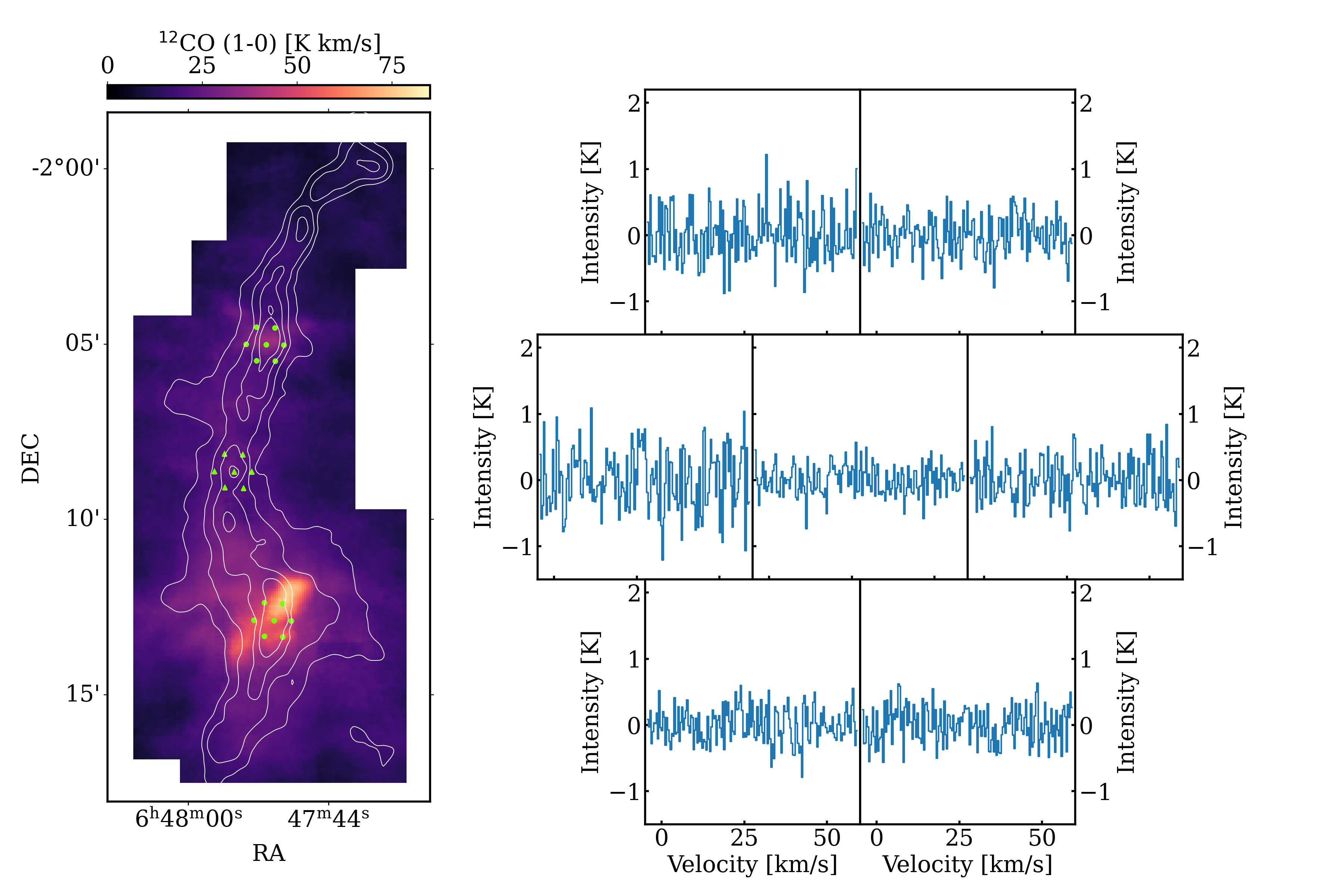}
\caption{(Left) A moment zero map of $\co$ (1-0) with the SOFIA pointing locations denoted as green symbols. (Right) The [CII] spectra of the individual pixels corresponding to the C14 pointing, seen as green triangles in the left panel, with the same arrangement as the pixel locations.}
\label{fig::app_CII}
\end{figure*}

Figure \ref{fig::app_CII} shows the location of the 3 SOFIA [CII] pointings, 7 pixels per pointing, on top of the $\co$ (1-0) moment zero for G214.5, as well as the individual spectra for the 7 pixels for the C14 pointing. We use the non-detection of [CII] in the pixels away from clump C11, to place an upper limit on the FUV in the vicinity of G214.5. For this purpose we use the predicted ratio of [CII]/CO (J=1-0) from the 2020 Wolfire/Kaufman PDR models included in PDR toolbox\footnote{https://dustem.astro.umd.edu}\citep{Kau06,Wol10,NeuWol16}. 

\begin{figure}
\centering
\includegraphics[width=0.8\linewidth, trim={2.1cm 0.5cm 0.5cm 0cm},clip]{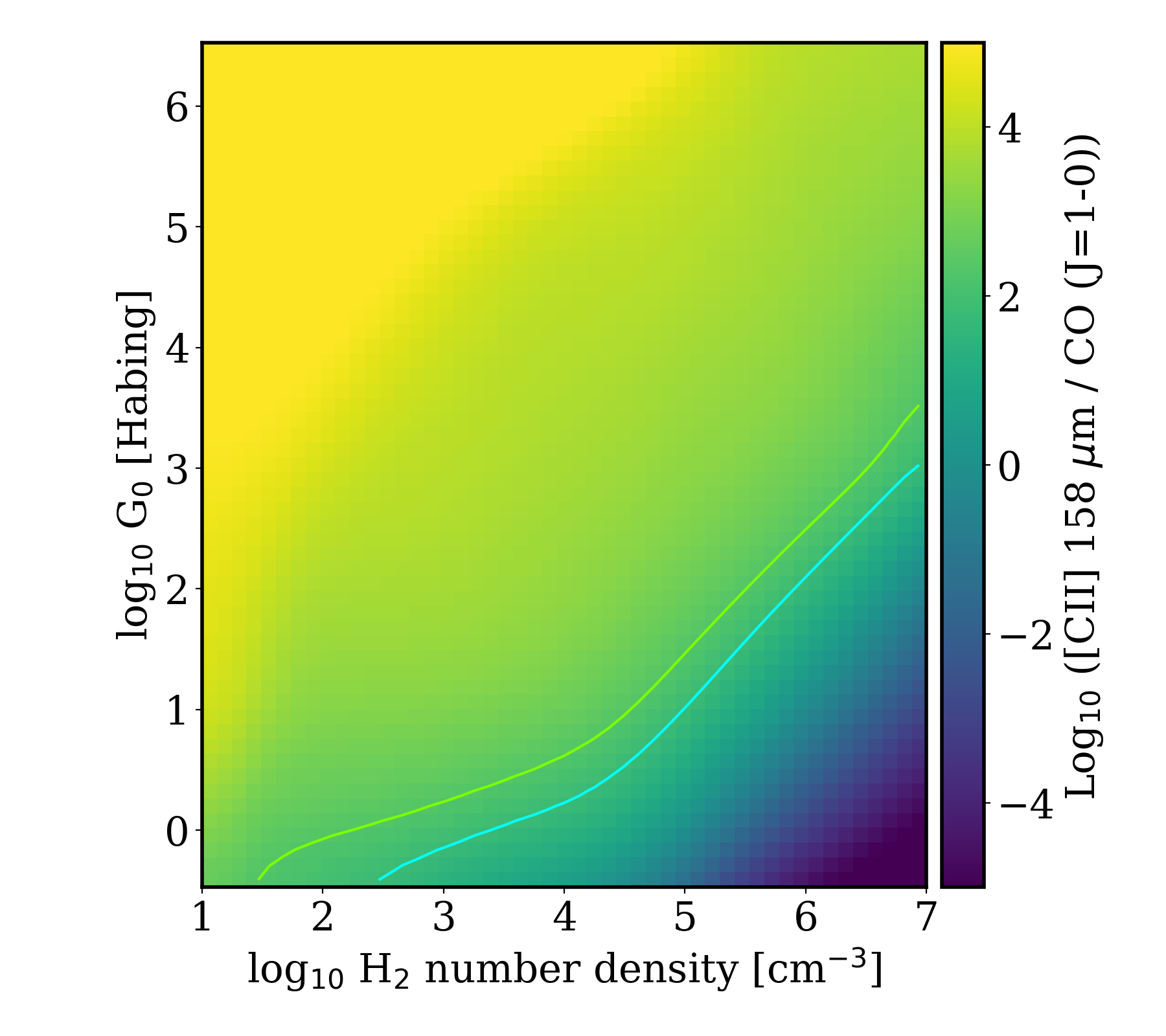}
\caption{The predicted [CII]/CO (J=1-0) emission from the 2020 models of the PDR toolbox \citep{Kau06,Wol10,NeuWol16} as a function of H$_2$ number density and FUV flux. The green and cyan contours denote the upper limit values of 207.9 and 65.8 respectively, corresponding to the median and minimum values of the [CII]/CO (J=1-0) ratio upper limits across the 20 non-detections.}
\label{fig::app_PDR}
\end{figure}

The noise level from each spectrum ranges from $\sim0.20 - 0.50$ K depending on the pixel and the pointing. Assuming a FWHM of 1 km/s for the [CII] line, using these noise levels results in 3$\sigma$ upper limits of $4.49\times10^{-6}$-$1.10\times10^{-5}$ erg/s/sr/cm$^{-2}$. Dividing these upper limits for the [CII] line by the observed $\co$ moment zero at those positions results in ratios between 65.8 and 380.7, with a median value of 207.9. The minimum and median value of this ratio may be seen in figure \ref{fig::app_PDR} as the two contours. Using volume density information found when modelling the CO freeze-out in section \ref{SSEC:FREEZE}, we constrain the number density to below $10^4$ cm$^{-3}$. With this constraint, it places upper limits for the FUV flux at $\sim 1.1$ and $\sim 2.5$ G$_0$.

\begin{figure}
\centering
\includegraphics[width=0.9\linewidth]{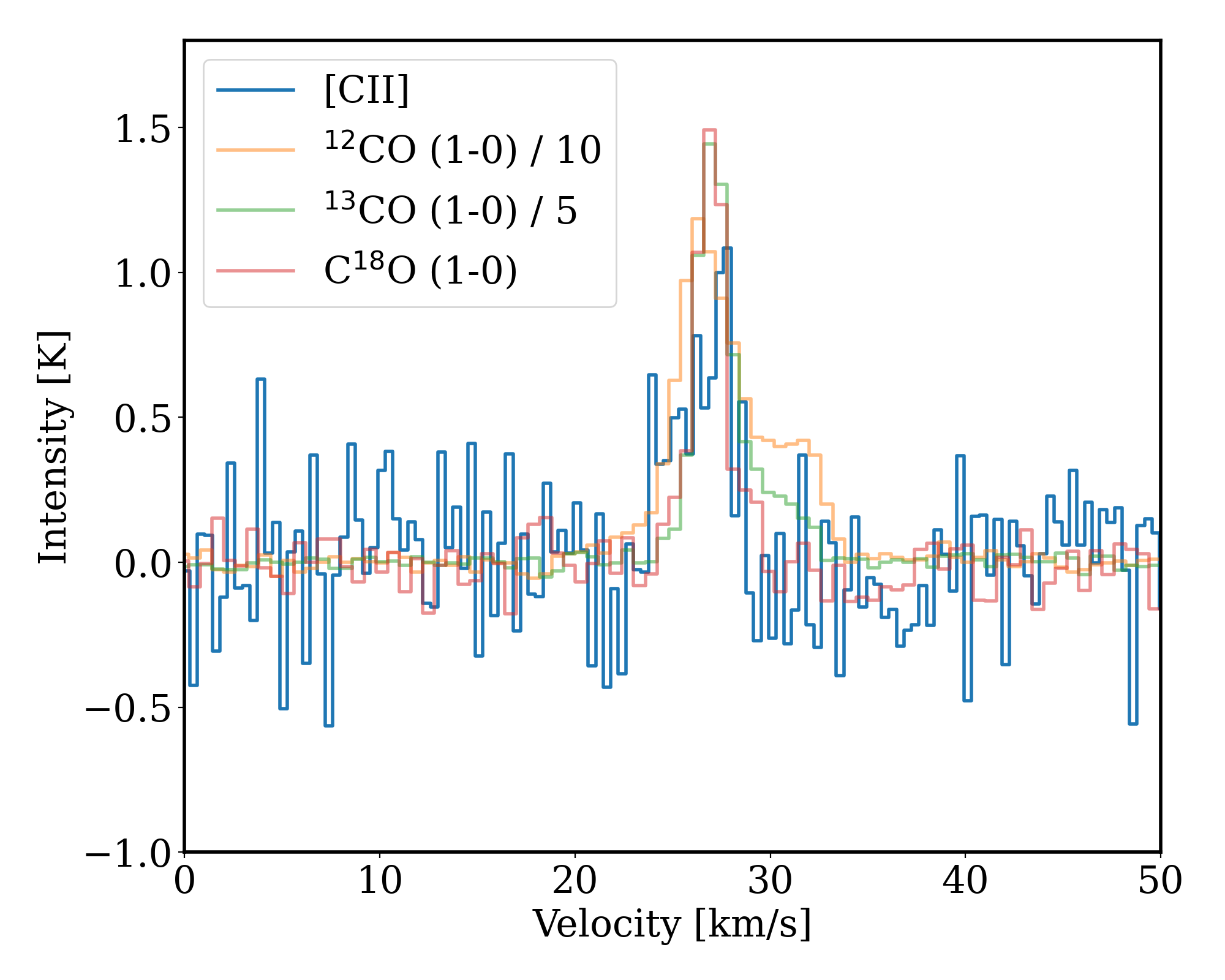}
\caption{The spectra at the centre of clump C11, (RA,DEC) = (06:47:50.2,-02:12:53): [CII] in blue, $\co$ in orange, $\coc$ in green and $\coo$ in red. The intensity of the $\co$ and $\coc$ spectra have been reduced by a factor of 10 and 5 respectively so that they may lie in the same range at the [CII] spectrum.}
\label{fig::app_detect}
\end{figure}

The single detection of [CII] towards the centre of clump C11 can be seen in figure \ref{fig::app_detect}, with the $\co$, $\coc$ and $\coo$ spectra at the same location. The detection is clear with a peak intensity of $\sim 1.1$ K, and lies at the same velocity as the CO isotopologues. 

\section{Monte Carlo error propagation to determine uncertainties}\label{APP:ERROR}%

\begin{figure*}
\centering
\includegraphics[width=0.9\linewidth]{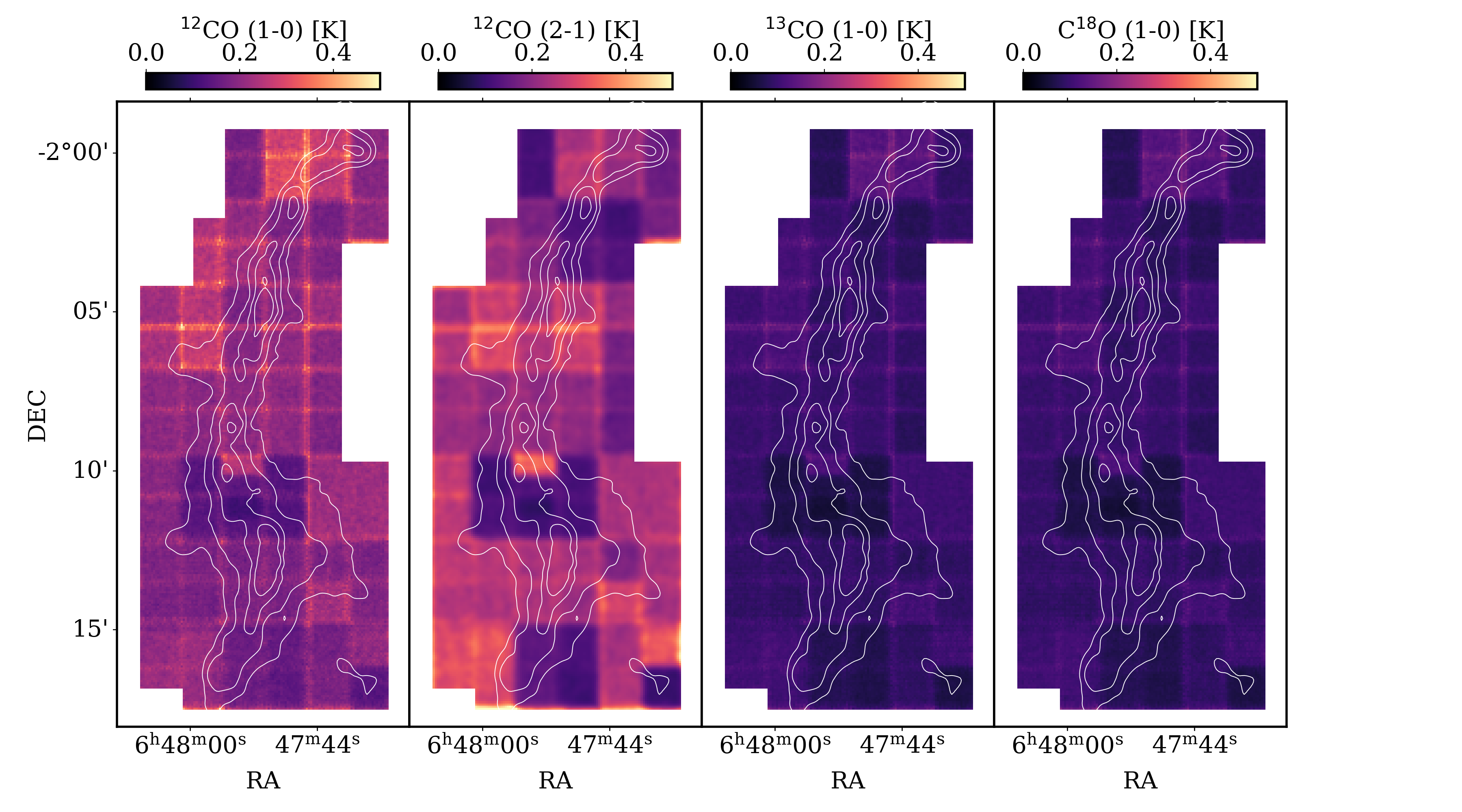}
\caption{Noise maps for, from left to right, $^{12}$CO (1-0), $^{12}$CO (2-1), $^{13}$CO (1-0), C$^{18}$O (1-0). The contours denote total column density at an A$_v$ of 2, 3, 5 and 8 magnitudes.}
\label{fig::app_noise}
\end{figure*}

To take into account the effects of noise in the observations on the calculated column densities and abundances we use a Monte Carlo error propagation method. This first requires knowledge of the pixel-by-pixel noise of the spectra; we determine this by calculating the standard deviation of the spectra intensity in velocity channels which are determined to be free of emission by the BTS moment-masking method described in section \ref{SEC:RES}. The resulting noise maps are shown in figure \ref{fig::app_noise}.

Using the pixel-by-pixel noise, we produce 10,000 realisations of the spectra per pixel where the random numbers are drawn from a normal distribution with a mean of the observed intensity in the velocity channel and the standard deviation is the pixel noise. Using these spectral realisations, we calculate the moment zero map, $\co$ excitation temperatures, $\coc$ and $\coo$ column densities and abundances, as well as the abundance ratio of $^{13}$CO over C$^{18}$O in the same manner as that done in the main body of the text. The standard deviation of these 10,000 realisations may then be used as a measure of the error on the calculated quantity. The maps of these errors, both as an absolute value and as a fraction of the quantity, can be seen in the supplementary online material. It should be noted that the propagated error calculated in this way is solely due to the noise in the spectra and does not include the systematic errors/biases which may be introduced due to assumptions in the analysis.

\section{Comparing the $\co$ and $\coc$ excitation temperatures}\label{APP:RADEX}%

To examine the assumption that the $\co$ and $\coc$ excitation temperatures are comparable we determine the $\coc$ excitation temperature using non-LTE RADEX calculations constrained by the APEX $\coc$ (3-2) and IRAM 30m $\coc$ (1-0) lines. We run a set of 4096 RADEX models using the LVG approximation with temperatures ranging from 5-20 K in steps of 1 K, densities ranging from $10^2$-$10^5$ cm$^{-3}$ in steps of 0.2 dex, and column densities of $\coc$ ranging from 10$^{14}$-10$^{17}$ cm$^{-2}$ in steps of 0.2 dex. The linewidth is taken to be 1 km/s, comparable to the observed linewidths of $\coc$.

\begin{figure*}
\centering
\includegraphics[width=0.9\linewidth, trim={6cm 1.5cm 5.2cm 0cm},clip]{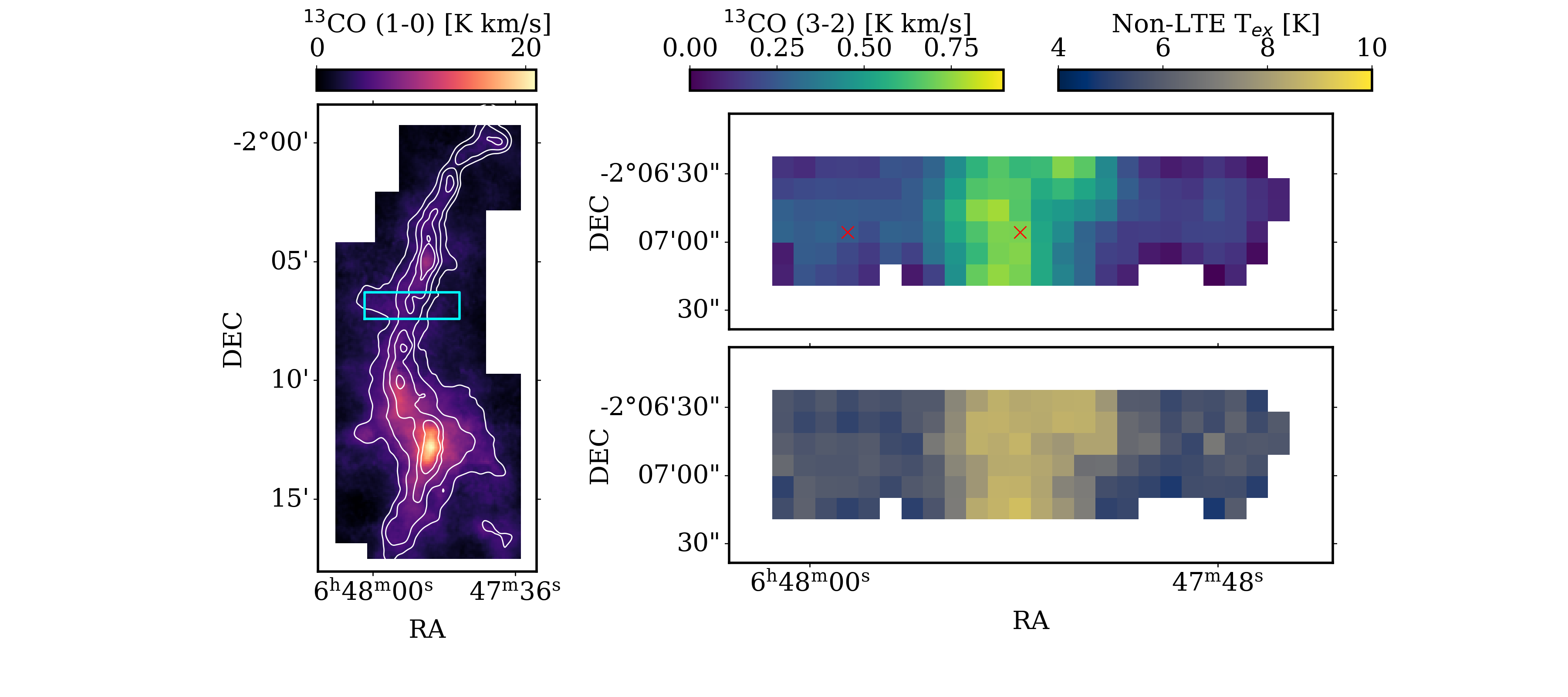}
\caption{(Left) A moment zero map of the $^{13}$CO (1-0) transition, the same as shown in figure \ref{fig::mom}. The cyan rectangle shows the region mapped by APEX. (Top right) A moment zero map of the $^{13}$CO (3-2) transition. The red crosses denote the locations of the spine and off-spine spectra shown in figure \ref{fig::app_radex_spec}. (Bottom right) A map showing the excitation temperature of $^{13}$CO determined using non-LTE RADEX calculations.}
\label{fig::app_radex_tex}
\end{figure*}

We calculate the moment zero map for the APEX $\coc$ (3-2) using BTS as done for the (1-0) line, as well as the error on the moment zero. The APEX data is smoothed to the common 23.5'' resolution before the moment zero is calculated. The moment zero map can be seen in the top right panel of figure \ref{fig::app_radex_tex}. To determine the excitation temperature we minimise the $\chi^2$ as defined as:
\begin{equation}
\chi^2 = \frac{(O_{rat} - M_{rat})^2}{E_{rat}^2},
\end{equation}
where $O_{rat}$ is the observed ratio of the (3-2) moment zero over the (1-0) moment one, $M_{rat}$ is the predicted ratio from the RADEX model, and $E_{rat}$ is the error on the observed ratio. The resulting excitation temperature map is shown in the lower right panel of figure \ref{fig::app_radex_tex}. As in the main body of this work, we utilise a Monte Carlo method to propagate the errors from the moment maps to the estimate of the excitation temperature. To do this we calculate 10,000 realisations of the moment zero values for the (3-2) and (1-0) lines, and for each realisation we minimise the above $\chi^2$ to determine 10,000 estimates of the $\coc$ excitation temperature, the standard deviation of which may be taken as the uncertainty in the excitation temperature. 

\begin{figure}
\centering
\includegraphics[width=0.9\linewidth]{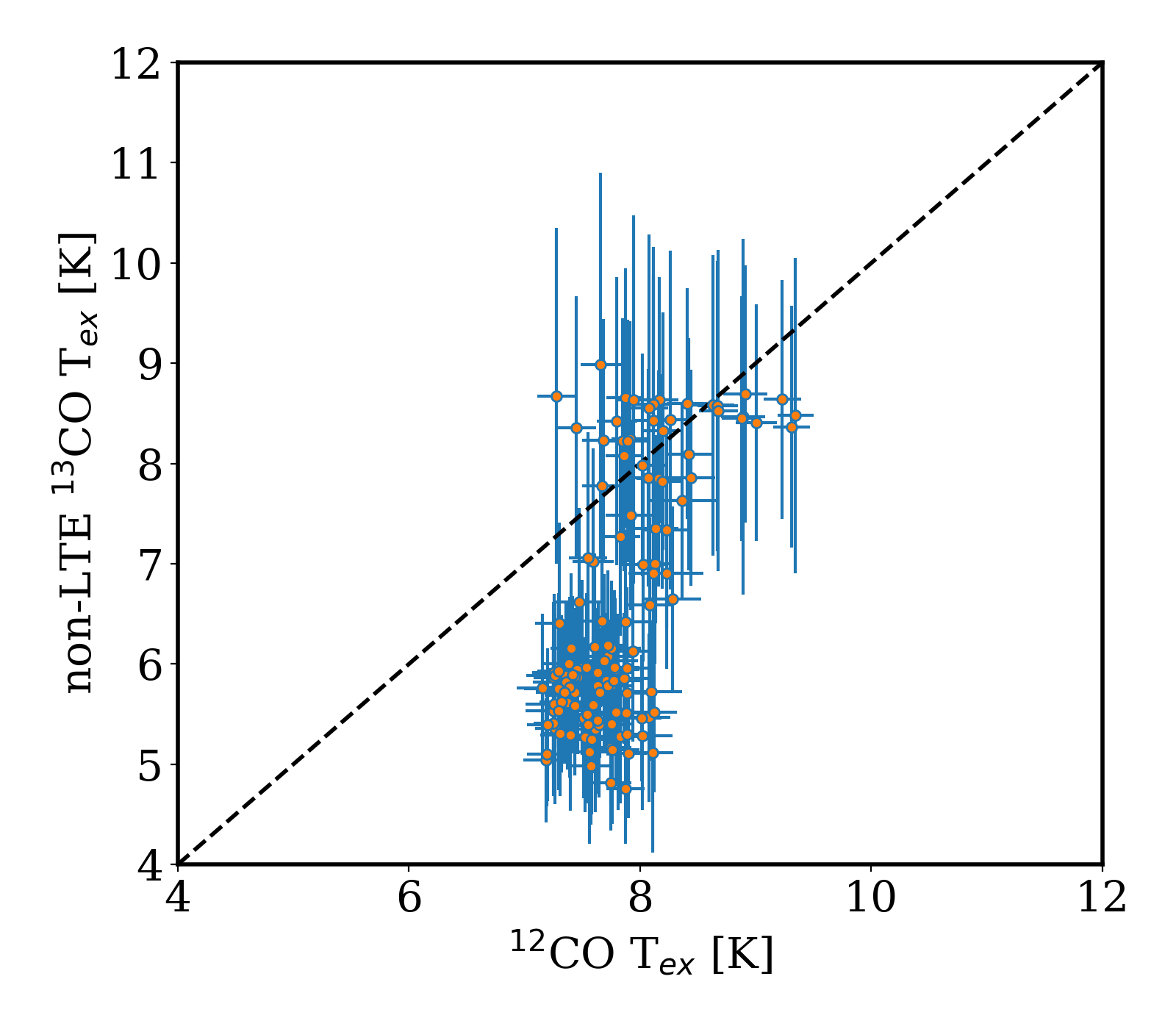}
\caption{A scatter plot showing the comparison of the $^{12}$CO excitation temperature calculated in section \ref{SEC:NCO} and the $^{13}$CO excitation temperature calculated using non-LTE RADEX calculations. The dashed, black line shows the 1:1 relationship.}
\label{fig::app_radex_tex_scatter}
\end{figure}

\begin{figure}
\centering
\includegraphics[width=0.8\linewidth, trim={0cm 3.2cm 0cm 3cm},clip]{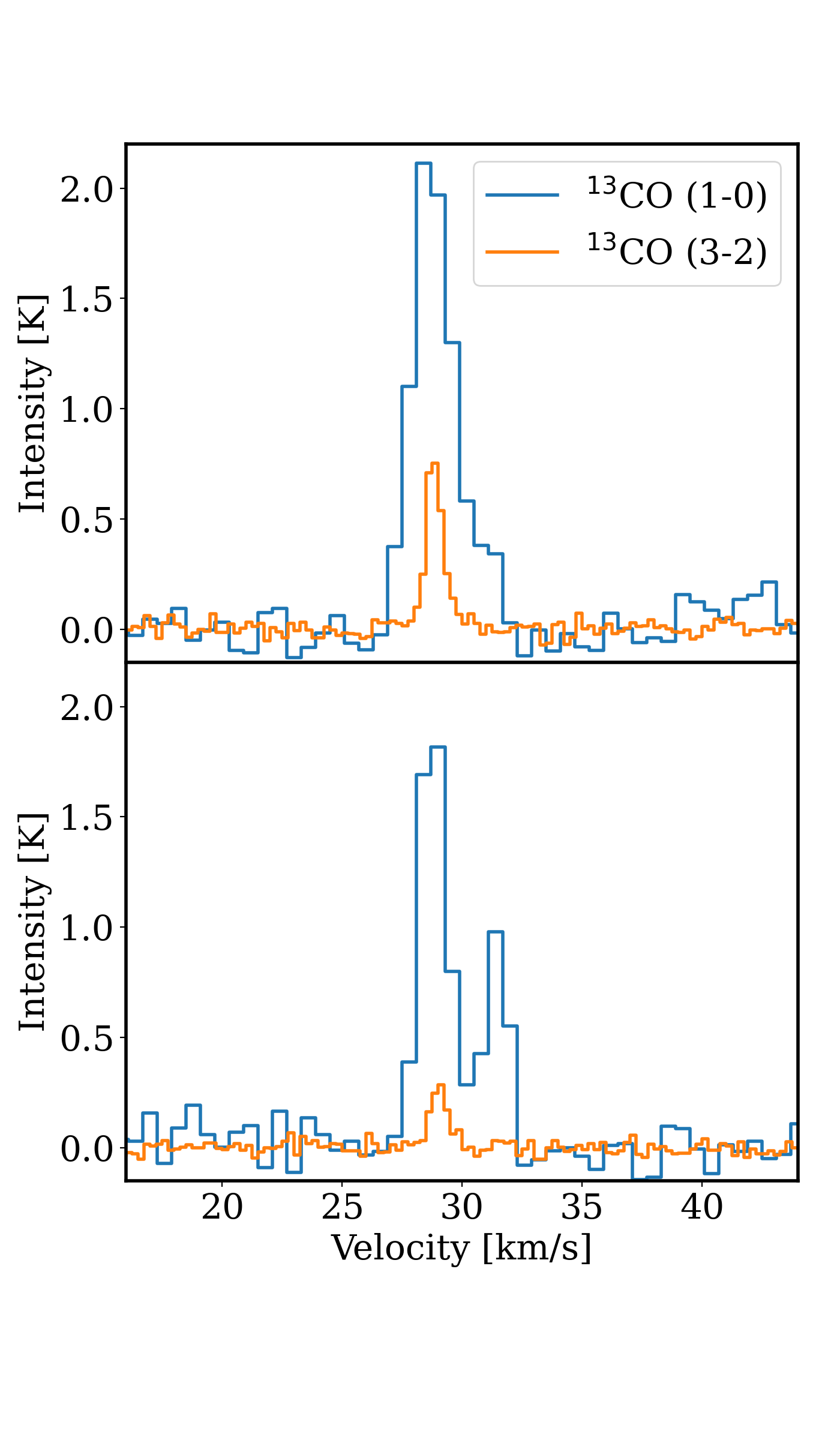}
\caption{$^{13}$CO spectra of the (1-0) transition (blue), and the (3-2) transition (orange), at the spine (top panel) and off-spine (bottom panel) locations shown as red crosses in \ref{fig::app_radex_tex}.}
\label{fig::app_radex_spec}
\end{figure}

In figure \ref{fig::app_radex_tex_scatter} we compare the excitation temperature calculated from $\co$ and $\coc$ and find that within errors most are consistent with each other. There is a tendency for the $\coc$ excitation temperature to be slightly lower than the $\co$ excitation temperature by around 0.5 K but this is minor compared to the 70$\%$ difference seen in more active regions \citep[e.g.][]{Rou21}. There is a group of pixels which lie around 2 K below the 1:1 line; however, these pixels also exhibit multiple velocity components in the $\coc$ (1-0) which are not seen in the (3-2) line (an example of a simple, single velocity component pixel and a multi-component pixel is shown in figure \ref{fig::app_radex_spec}). For such pixels the moment zero ratio of (3-2) over (1-0) will be artificially lowered, thus resulting in a lower excitation temperature than is associated with the shared velocity component. As such we conclude that the $\co$ derived and $\coc$ derived excitation temperatures are broadly consistent with each other and the $\co$ excitation temperature may be used for $\coc$ and $\coo$ column density estimates. 

\end{appendices}

\label{lastpage}

\end{document}